\documentclass[
  10pt,
  twocolumn,
  english,
  aps,
  pra,
  longbibliography,
  superscriptaddress,
  floatfix
]{revtex4-1}
\usepackage{amsmath}
\usepackage{amssymb}
\usepackage{mathtools}
\usepackage{graphicx}
\usepackage{physics}
\usepackage[dvipsnames]{xcolor}
\usepackage{amsthm}
\usepackage[english]{babel}
\usepackage{quantikz}
\usepackage{multirow}

\mathtoolsset{showonlyrefs=true}

\usepackage{silence}
\usepackage[colorlinks=true,urlcolor=blue,citecolor=blue,linkcolor=blue]{hyperref}

\newcommand*{\brakets}[1]{\langle #1 \rangle}

\newcommand{\calV}{\mathcal{V}}

\newcommand{\calE}{\mathcal{E}}

\begin{document}

\title{Efficient Fourier-Based Linear Combination of Unitaries\\and Applications in Quantum Optimization}
\author{Almudena Carrera Vazquez}
\affiliation{IBM Research, Saeumerstrasse 4, CH-8803 Rueschlikon, Switzerland}
\author{Daniel J.~Egger}
\affiliation{IBM Research, Saeumerstrasse 4, CH-8803 Rueschlikon, Switzerland}
\author{Stefan Woerner}
\email{wor@zurich.ibm.com}
\affiliation{IBM Research, Saeumerstrasse 4, CH-8803 Rueschlikon, Switzerland}
\date{\today{}}

\begin{abstract}
We investigate ancilla-free linear combination of unitaries (LCU) as a framework for approximating complex quantum circuits. This is particularly effective for quantum optimization algorithms, where candidate solutions can be evaluated classically and the task is to sample high-quality bitstrings rather than reproduce the full output distribution. We show that Fourier-based LCU constructions efficiently decompose broad classes of diagonal and non-diagonal unitaries, replacing highly connected qubit interactions with single-qubit gate layers or significantly simpler structures at the cost of a polynomial sampling overhead.
Applied to algorithms such as QAOA, this yields efficient, hardware-friendly decompositions of, for instance, cardinality-constraint penalties and the fully connected XY-mixer, while maintaining rigorous performance guarantees compared to fully coherent implementations. Furthermore, we establish a formal connection between Fourier-based quantum penalties and Lagrangian relaxation, offering a unified perspective on constraint handling.
We validate our approach using exact statevector simulations of 12-qubit circuits and large-scale experiments on 106 superconducting qubits. Our results illustrate how approximate sampling via an LCU systematically trades circuit complexity for sampling overhead, extending the practical reach of near-term quantum optimization.
\end{abstract}

\maketitle


\section{Introduction}

Quantum optimization is one of the most prominent application areas for near- and intermediate-term quantum computers \cite{Abbas_2024, koch_2025_quantum_optimization_benchmark_library}. Many practically relevant problems, ranging from logistics and scheduling to finance, energy systems, and biology, can be formulated as optimization problems. Quantum algorithms such as the Quantum Approximate Optimization Algorithm (QAOA) provide a natural approach to such problems \cite{farhi_2014_qaoa, hadfield_2019_xymixer, wang_2020_xymixer}.

A central challenge, however, is that the circuits arising from the mathematical structure of an optimization problem, particularly in the presence of constraints, are often too complex to be implemented efficiently on current or near-term hardware. Direct implementations typically require long-range gates, a large number of two-qubit gates, and substantial circuit depth, often exceeding hardware capabilities \cite{weidenfeller_2022_scaling_qaoa_circuits}.

This motivates approaches that avoid implementing the full target circuit. Possible strategies include approximating circuits to trade solution quality for reduced circuit complexity \cite{dragoi_2026_hubo}, or replacing complex circuits with a larger number of simpler circuits at the cost of increased sampling overhead, as in circuit cutting or multi-product formulas \cite{Peng_2020_large_circuits_small_qc, Mitarai_2021_virtual_two_qubit_gate, Perlin_2021_circuit_cutting_mlft, Tang_2021_CutQC, Piveteau_2024_circuit_knitting_classical_communication, Lowe_2023_fast_circuit_cutting, Brenner_2025_optimal_wire_cutting, Schmitt_2025, Harrow_2025_optimal_circuit_cuts, Vazquez_2022, Carrera_Vazquez_2023}. 

In this work, we introduce a Fourier-based Linear Combination of Unitaries (LCU) framework for quantum circuits arising, among others, in quantum optimization algorithms. In contrast to circuit-cutting, which typically decomposes circuits locally, our approach performs a global decomposition. The key idea is to exploit Fourier structure to express unitaries that are difficult to implement coherently as linear combinations of unitaries that are easier to implement, while analytically controlling the resulting approximation and sampling overhead. These LCU decompositions can be realized via Quasi-probability Decompositions (QPD) over physical basis operations, incurring a sampling overhead \cite{Temme_2017_error_mitigation, Endo_2018_practical_qem, Piveteau2022QPD, van_den_Berg_2023_sparse_pauli_lindblad_pec}.

The framework is especially natural in the context of QAOA. We show how Fourier-based LCU decompositions can replace QAOA cost and mixer operators that would otherwise require all-to-all qubit connectivity with ensembles of significantly simpler circuits involving only single-qubit gates.
Their outcomes are recombined classically, often incurring only a polynomial sampling overhead. The framework is also applicable to other sampling-based algorithms in which sampled bitstrings are evaluated classically, such as Sampling-based Quantum Diagonalization (SQD) and its variants \cite{Robledo_Moreno_2025, yu2025quantumcentricalgorithmsamplebasedkrylov, piccinelli2026quantumchemistryprovableconvergence}.

We test our framework using QAOA on the \emph{densest $k$-subgraph} problem \cite{feige_2001_dense_k_subgraph}. First, we analyze its performance and validate the theory using illustrative statevector simulations on a 12-node instance. We then demonstrate scalability on a 106-node instance, executed on 106 qubits of an IBM Quantum computer, illustrating the practical applicability of the approach.

The remainder of this paper is organized as follows. Sec.~\ref{sec:lcu} introduces the LCU framework, including the ancilla-free variant. Sec.~\ref{sec:fourier_diagonal} develops the Fourier-based LCU for diagonal unitaries. Sec.~\ref{sec:fourier-nondiagonal} generalizes the construction to non-diagonal permutation-invariant unitaries, such as the XY-mixer. Sec.~\ref{sec:qaoa_lcu} discusses the application to QAOA, highlighting connections between coherent penalty formulations and Lagrangian relaxations. Sec.~\ref{sec:demonstrations} presents numerical simulations and hardware results, and Sec.~\ref{sec:conclusion} concludes with an outlook.

\section{Linear Combination of Unitaries
\label{sec:lcu}}

In this section, we introduce the LCU framework and describe how to construct a QPD implementation of a given LCU using an ancilla qubit. 
We then show how ancilla-free LCU methods can be employed when the goal is to sample relevant bitstrings with high probability, rather than to reproduce the exact target distribution. 
For further details on LCU and the construction presented here, we refer to the related literature \cite{childs_2012_lcu,berry_2015_lcu_taylor,Faehrmann_2022,Chakraborty2024implementingany,Schmitt_2025}.

\subsection{LCU and channel decomposition}
Consider an $n$-qubit unitary $U$ expressed as a linear combination of unitaries
\begin{eqnarray}
    U &=& \sum_{j=0}^m c_j V_j,
\end{eqnarray}
where $c_j = |c_j| e^{i \phi_j} \in \mathbb{C}$ and each $V_j$ is unitary.
The induced quantum channel $\mathcal{U}(\rho) = U \rho U^{\dagger}$ can be expanded as
\begin{eqnarray}
    \mathcal{U}(\rho) = 
        \sum_{j=0}^m |c_j|^2 V_j \rho V_j^{\dagger} 
        + \sum_{k<j} 2 |c_j| |c_k| \mathcal{W}_{jk}(\rho), \nonumber
\end{eqnarray}
where
\begin{eqnarray}
    \mathcal{W}_{jk}(\rho) &=& 
        \frac{1}{2}\left( e^{i \phi_{jk}} V_j \rho V_k^{\dagger} + e^{-i \phi_{jk}} V_k \rho V_j^{\dagger} \right),
\end{eqnarray}
and $\phi_{jk} = \phi_j - \phi_k$.

Each cross term $\mathcal{W}_{jk}(\rho)$ can be implemented using two Kraus operators
\begin{eqnarray}
K_{jk}^{\pm} &=& \frac{1}{2}\left(e^{i \phi_{jk}/2} V_j \pm e^{-i \phi_{jk}/2} V_k\right),
\end{eqnarray}
leading to the identity
\begin{eqnarray}
    \Phi_{jk}^+(\rho) - \Phi_{jk}^-(\rho) = \mathcal{W}_{jk}(\rho),
\end{eqnarray}
where $\Phi_{jk}^{\pm}(\rho) = K_{jk}^{\pm} \rho K_{jk}^{\pm \dagger}$.

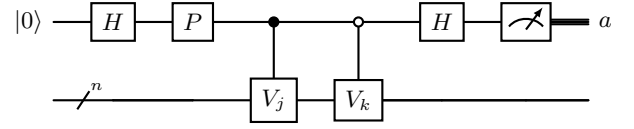
\begin{figure}
\[
\begin{quantikz}
\lstick{$\ket{0}$}   & \gate{H} & \gate{P} & \ctrl{1} & \octrl{1} & \gate{H} & \meter{} & \cw \rstick{$a$} \\
 & \qwbundle{n}      & \qw      & \gate{V_j} & \gate{V_k} & \qw      & \qw      & \qw
\end{quantikz}
\]
\caption{Quantum circuit for implementing $\Phi_{jk}^+(\rho) - \Phi_{jk}^-(\rho)$, where $P$ denotes a phase gate applying $\phi_{jk}$.}
\label{fig:lcu_circuit}
\end{figure}

This yields a QPD of the full channel
\begin{eqnarray}
    \mathcal{U}(\rho) &=& 
        \sum_{j=0}^m |c_j|^2 V_j \rho V_j^{\dagger} + \\
        &&\sum_{k<j} 2 |c_j| |c_k| \left( \Phi_{jk}^+(\rho) - \Phi_{jk}^-(\rho) \right).
\end{eqnarray}
A standard implementation realizes $( \Phi_{jk}^+ - \Phi_{jk}^- )$ using an ancilla qubit together with controlled applications of $V_j$ and $V_k$, as illustrated in Fig.~\ref{fig:lcu_circuit}. 
The ancilla is measured, producing an outcome $a\in\{0,1\}$ that determines a sign factor $(-1)^a$.

Consequently, expectation values of observables can be estimated directly from samples of this randomized procedure.
For an observable $O$, we obtain
\begin{eqnarray}
    \text{tr}(O\mathcal{U}(\rho)) &=& \sum_{j=0}^m |c_j|^2 \text{tr}(O V_j \rho V_j^{\dagger}) + \\
    && \sum_{k<j} 2 |c_j| |c_k| \mathbb{E}_a \left[ (-1)^a \text{tr}(O \rho_a) \right],    
\end{eqnarray}
where $\rho_a$ denotes the post-measurement state conditioned on obtaining outcome $a$.

The total QPD cost is
\begin{equation*}
    \Gamma
    = \sum_{j=0}^m |c_j|^2 + \sum_{k < j} 2|c_j| |c_k|
    =  \| c \|_1^2,
\end{equation*}
and estimating an observable up to additive error $\epsilon > 0$ requires $\mathcal{O}(\Gamma^2 / \epsilon^2 )$ samples~\cite{Piveteau2022QPD}.

\subsection{Ancilla-free LCU for sampling\label{subsec:ancilla_free_lcu}}
In many applications, the goal is not to estimate expectation values with high precision, but rather to sample \emph{good} bitstrings with high probability.
In this setting, the QPD can be simplified, reducing both sampling overhead and circuit complexity \cite{Chakraborty2024implementingany, Barron_2024}.

Consider measuring the observable $O = \proj{x}$ and denote the resulting probability of measuring $\ket{x}$ by 
\begin{equation*}
    p_x \coloneqq \mathrm{tr}(\proj{x}\mathcal{U}(\rho)).
\end{equation*}
Dropping the sign factor $(-1)^a$ from the QPD can only increase the contribution of each term, leading to the upper bound
\begin{eqnarray}
    p_x &\leq& \sum_{j=0}^m |c_j|^2 \text{tr}(\proj{x} V_j \rho V_j^{\dagger}) \\
    && + \sum_{k<j} 2 |c_j| |c_k| \mathbb{E}_a \left[ \text{tr}(\proj{x} \rho_a) \right].
\end{eqnarray}
If the measurement outcome is not used, the ancilla measurement and its surrounding gates can be omitted without affecting the sampling probabilities. 
In particular, the final Hadamard gate in Fig.~\ref{fig:lcu_circuit} becomes irrelevant, and the phase gate can also be removed since it commutes with the controlled operations. 

This observation allows us to replace the ancilla qubit by classical randomness, where $V_j$ and $V_k$ are applied uniformly at random, resulting in
\begin{eqnarray}
    p_x &\leq& \sum_{j=0}^m |c_j|^2 \text{tr}(\proj{x} V_j \rho V_j^{\dagger}) + \\
    && \sum_{k<j} |c_j| |c_k|  \left[ \text{tr}(\proj{x} V_j\rho V_j^{\dagger}) + \text{tr}(\proj{x} V_k\rho V_k^{\dagger}) \right]    \\
    &=& \Gamma \sum_{j=0}^m q_j \text{tr}(\proj{x} V_j \rho V_j^{\dagger}) \eqqcolon \Gamma \widetilde{p}_x,
\end{eqnarray}
where $q_j = |c_j| / \sqrt{\Gamma}$.
Thus, we obtain the bound 
\[
\widetilde{p}_x \geq \frac{p_x}{\Gamma}.
\]

In other words, instead of implementing the exact QPD, we may sample unitaries $V_j$ according to $q_j$ and compensate by taking $\Gamma$ times more samples. 
This preserves the probability of sampling \emph{good} bitstrings up to the factor $\Gamma$.

Importantly, this removes the need for an ancilla qubit and controlled operations, substantially simplifying the circuit implementation. 
Moreover, as shown in \cite{Barron_2024}, the Conditional Value at Risk (CVaR) at level $\alpha = 1/\Gamma$ yields provable bounds on expectation values. 
CVaR is defined as the expectation value over a specified tail of a distribution, determined by the risk level $\alpha$.
Specifically, letting $X$ and $\widetilde{X}$ be sampled from $p_x$ and $\widetilde{p}_x$, respectively, we have 
\begin{eqnarray}
\text{CVaR}_{1/\Gamma}(f(\widetilde{X})) \leq \mathbb{E}[f(X)] \leq \overline{\text{CVaR}}_{1/\Gamma}(f(\widetilde{X}))
\end{eqnarray}
for any function $f: \{0, 1\}^n \rightarrow \mathbb{R}$. Here, CVaR and $\overline{\text{CVaR}}$ denote the lower- and upper-tail CVaR, respectively. In the remainder of the paper, we simply write CVaR, since it is clear from the context whether we refer to the lower- or upper-tail version. 
In the following, we leverage these insights to construct efficient approximations of operations that would otherwise be prohibitively costly to implement exactly on near-term quantum hardware.

\section{Fourier-based LCU for diagonal unitaries
\label{sec:fourier_diagonal}}

The approach introduced in Sec.~\ref{sec:lcu} is particularly effective when a target unitary admits an LCU with small cost $\Gamma$.
In this section, we present a construction based on the discrete Fourier transform that yields efficient decompositions for a broad class of diagonal unitaries.
Efficient implementations of diagonal unitaries without ancillas have been studied previously, for example in~\cite{Welch_2014}, but can still lead to very complex circuits with nonlocal interactions.
In contrast, our Fourier-based LCU construction replaces these operations by ensembles of much simpler circuits, often involving only single-qubit gate layers, at the cost of a controlled sampling overhead.

Consider a parametrized family of diagonal unitaries defined by a function $g: \{0, 1\}^n \rightarrow \{0, \ldots, m\}$,
\begin{eqnarray}
    V(\theta) &=& \sum_{x \in \{0, 1\}^n} e^{i \theta g(x)} \proj{x},
\end{eqnarray}
which we assume can be implemented efficiently on a quantum computer.
We aim to realize a target unitary of the form
\begin{eqnarray}
    U_f(\gamma) &=& \sum_{x \in \{0, 1\}^n} e^{-i \gamma f(g(x))} \proj{x},
\end{eqnarray}
for an arbitrary function $f: \{0, \ldots, m\} \rightarrow \mathbb{R}$.

Define $\theta_j = 2\pi j / (m+1)$ and coefficients
\begin{eqnarray}
    c_j &=& \frac{1}{m+1}\sum_{k=0}^m e^{-i \gamma f(k)} e^{-i \theta_j k},
\end{eqnarray}
i.e., the discrete Fourier transform of $\{e^{-i \gamma f(k)}\}_{k=0}^m$.
Applying the inverse discrete Fourier transform, we obtain
\begin{eqnarray}
    e^{-i \gamma f(g(x))} &=& \sum_{j=0}^m c_j e^{i \theta_j g(x)},
\end{eqnarray}
which directly yields the LCU  
\begin{eqnarray}
    U_f(\gamma) &=& \sum_{j=0}^m c_j V(\theta_j).
\end{eqnarray}
Since the coefficients $c_j$ correspond to the Fourier transform of a unit-modulus function, Parseval's identity implies $\|c\|_2 = 1$~\cite{oppenheim1999signals}, and hence $\|c\|_1 \leq \sqrt{m+1}$, giving $\Gamma \leq m+1$.

A particularly relevant case is given by permutation-invariant diagonal unitaries, where $f$ depends only on the Hamming weight of $x \in \{0, 1\}^n$.
In this case, $g(x) = \mathbf{1}^T x$ and $m = n$, where $\mathbf{1}$ denotes the vector of all ones.
The corresponding unitaries take the form
\begin{eqnarray}
    V(\theta) &=& \bigotimes_{i=1}^n R_Z(\theta)
    = \bigotimes_{i=1}^n e^{-i \theta Z / 2} \\
    &=& \sum_{x \in \{0, 1\}^n} e^{-i \theta (n - 2 (\mathbf{1}^Tx)) / 2} \proj{x} \\
    &=& e^{-i \theta n/2} \sum_{x \in \{0, 1\}^n} e^{i \theta (\mathbf{1}^Tx)} \proj{x},
\end{eqnarray}
where we factor out the global phase in the last step. 
Thus, up to this irrelevant global phase, $V(\theta)$ realizes the unitary rotation corresponding to $g(x) = \mathbf{1}^T x$ using only a single layer of single-qubit $R_Z$ gates.

This is particularly useful in QAOA, where objective functions and constraints of binary combinatorial optimization problems are typically encoded in diagonal unitaries. 
For example, the permutation-invariant quadratic penalty term $f(g(x)) = (\mathbf{1}^T x - b)^2$ for decision variables $x \in \{0, 1\}^n$ and a target cardinality $b \in \{0, \ldots, n\}$, generally requires $\mathcal{O}(n^2)$ $R_{ZZ}$ gates and all-to-all connectivity. 
Using the Fourier-based LCU together with the exact QPD reduces this requirement to a star topology on $n+1$ qubits with the ancilla qubit at the center.
In contrast, the approximate sampling approach introduced in Sec.~\ref{subsec:ancilla_free_lcu} removes the need for ancillas and controlled operations altogether, reducing the implementation to a single layer of $R_Z$ gates.
Allowing more ancillas and interleaving ancilla and state qubits on a line also enables implementing the exact QPD to be implemented via mid-circuit measurements and feed-forward operations in constant depth \cite{Baeumer_2025}.

The same construction extends to more complex functions, such as $f(g(x)) = \mathbb{I}(\mathbf{1}^Tx \in [l, u])$, where $\mathbb{I}$ denotes an indicator function, enabling efficient implementations of inequality constraints on the Hamming weight.
Standard approaches, by contrast, generally require the introduction of slack variables together with all-to-all entangling gates or other highly nonlocal and complex circuit constructions.

The presented approach extends naturally to block-permutation-invariant diagonal unitaries. 
In this setting, we consider registers of qubits $\ket{x_1}\ldots\ket{x_b}$ and functions of the form $f(\mathbf{1}^T x_1, \ldots, \mathbf{1}^T x_b)$. 
For a fixed number of qubits per block, the resulting decomposition cost scales exponentially in the number of blocks.

More generally, suppose a diagonal unitary corresponding to a function $g: \{0, 1\}^n \rightarrow \{0, \ldots, m\}$, with $m = \text{poly}(n)$, admits an efficient implementation. This includes, for example, integer-valued or suitably discretized functions of the form $g(x) = x^T A x$ for some matrix $A$, implemented via $R_{ZZ}$ gates.
The Fourier construction then provides an efficient LCU for arbitrary compositions $f \circ g$, with $f: \{0, \ldots, m\} \rightarrow \mathbb{R}$.
This enables the implementation of penalty terms for equality and inequality constraints of the form $l \leq x^T A x \leq u$, where $l, u \in \{0, \ldots, m\}$ with $l < u$.
Standard approaches typically require the introduction of slack variables and $\mathcal{O}(n^4)$ $R_{ZZ}$ gates.
In contrast, our method relies solely on circuits implementing $V(\theta)$, at the cost of a sampling overhead of $m+1$.
This construction naturally extends to arbitrary polynomial functions over $\{0, 1\}^n$, as well as more general functions admitting a sparse Fourier representation, thereby substantially broadening the class of efficiently approximable diagonal unitaries.

\section{Fourier-based LCU for non-diagonal permutation-invariant unitaries
\label{sec:fourier-nondiagonal}}

We now extend the Fourier-based LCU construction from diagonal to general permutation-invariant unitaries.
While these are no longer diagonal, their symmetry imposes a structured block decomposition. 
Our strategy exploits this structure by expressing the unitary as a linear combination of collective single-qubit rotations, using a Fourier expansion over the group $SU(2)$.

Let $S_n$ denote the symmetric group on $n$ elements.
Each permutation $\pi\in S_n$ acts on the $n$-qubit Hilbert space $(\mathbb{C}^2)^{\otimes n}$ by permuting tensor factors via
\begin{equation*}
P_\pi\left(\ket{\psi_1}\otimes\cdots\otimes\ket{\psi_n}\right)=\ket{\psi_{\pi^{-1}(1)}}\otimes\cdots\otimes\ket{\psi_{\pi^{-1}(n)}}.
\end{equation*}
A unitary $U$ acting on $(\mathbb{C}^2)^{\otimes n}$ is called \textit{permutation-invariant} if it commutes with every permutation operator, i.e.
\begin{equation*}
    [P_\pi,U]=0\quad\text{for all } \pi\in S_n.
\end{equation*}

The \(n\)-qubit Hilbert space admits the total-spin decomposition
\[
(\mathbb C^2)^{\otimes n}
\simeq
\bigoplus_{j\in\mathcal J_n}
\mathbb C^{2j+1}\otimes \mathbb C^{m_j},
\]
where
\[
\mathcal J_n=
\left\{\frac n2,\frac n2-1,\dots,j_{\min}\right\},
\]
with \(j_{\min}=0\) for even \(n\) and \(j_{\min}=1/2\) for odd \(n\).
The multiplicities are given by $m_j=\binom{n}{\frac{n}{2}-j}-\binom{n}{\frac{n}{2}-j-1}$.
A detailed derivation of this decomposition is provided in Appendix~\ref{appendix:hilbert_decomposition}.

This decomposition is a consequence of the Schur--Weyl duality~\cite{goodman2009symmetry}, which states that the Hilbert space decomposes into irreducible representations of $SU(2)$ (spin-$j$ sectors) and $S_n$ (multiplicity spaces), with these actions commuting. Consequently, every permutation-invariant $n$-qubit unitary has the form
\[
U=
\bigoplus_{j\in\mathcal J_n}
U_j\otimes I_{m_j},
\]
where \(U_j\in U(2j+1)\). 
Thus, \(U\) acts non-trivially only on the spin-\(j\) spaces.

To construct such unitaries via an LCU, we seek a family of simpler unitaries that share this block structure.
We use collective single-qubit rotations as the LCU basis:
\[
V_g:=R(g)^{\otimes n},
\qquad g\in SU(2),
\]
where \(R(g)\) denotes the corresponding single-qubit representation of $g$, i.e., a generic single-qubit rotation. 
Any such rotation can be written in Euler form as
\[
R(\alpha,\vartheta,\chi)
=
e^{-i\alpha Z/2}
e^{-i\vartheta Y/2}
e^{-i\chi Z/2},
\]
so that \(R(g)\) can be implemented using single-qubit \(R_Z\) and \(R_Y\) rotations. 
These operators are permutation-invariant and decompose according to the same spin sector structure as
\[
R(g)^{\otimes n}
=
\bigoplus_{j\in\mathcal{J}_n}D^j(g)\otimes I_{m_j},
\]
where \(D^j(g)\) is the $(2j+1)$-dimensional irreducible representation of $SU(2)$ acting on the spin-\(j\) sector.
By Schur--Weyl duality with the double-commutant theorem~\cite{goodman2009symmetry}, the complex algebra generated by the collective rotations $R(g)^{\otimes n}$ coincides with the full commutant of $\{P_\pi:\pi\in S_n\}$.
In particular, on each spin-$j$ sector the matrices $D^j(g)$ generate all linear operators on $\mathbb{C}^{2j+1}$.

To express $U$ in this basis, we generalize the discrete Fourier decomposition used in the diagonal case to a continuous Fourier expansion over the compact group $SU(2)$, where sums over a finite set are replaced by integrals with respect to the normalized Haar measure $d\mu(g)$.
Accordingly, we seek a continuous LCU decomposition of the form
\[
U
=
\int_{SU(2)}
a_U(g)\, R(g)^{\otimes n}\, d\mu(g),
\]
for a suitable coefficient function \(a_U(g)\).
As shown in Appendix~\ref{appendix:fourier_in_groups}, these coefficients are given by
\[
a_U(g)
:=
\sum_{j\in\mathcal J_n}
(2j+1)\,
\operatorname{Tr}\!\left[
U_jD^j(g)^\dagger
\right].
\]

We now convert the continuous LCU into a form suitable for probabilistic implementation.
Define the normalization factor
\[
\alpha_U
:=
\int_{SU(2)}
|a_U(g)|\,d\mu(g).
\]
Writing the coefficient function in polar form as
\[
a_U(g)=|a_U(g)|e^{i\psi(g)},
\]
we obtain a probability density function
\[
q_U(g)=\frac{|a_U(g)|}{\alpha_U}.
\]
Then, the unitary $U$ can be expressed as
\[
U
=
\alpha_U
\mathbb E_{g\sim q_U}
\left[
e^{i\psi(g)}R(g)^{\otimes n}
\right].
\]
The associated channel QPD implementation has cost $\Gamma_U=\alpha_U^2$.
Operationally, the LCU-to-channel conversion proceeds by sampling independent group elements \(g,h\sim q_U\) and implementing the corresponding two-outcome instrument as in the discrete case.

In Appendix~\ref{appendix:bound_on_gamma}, we show that
\begin{equation*}
    \Gamma_U\leq \frac{(n+1)(n+2)(n+3)}{6}=O(n^3).
\end{equation*}
Thus, every permutation-invariant \(n\)-qubit unitary admits an exact LCU decomposition over collective rotations \(R(g)^{\otimes n}\), with induced channel-QPD cost at most \(O(n^3)\). 
If the ancilla instrument is replaced by classical branch sampling, the resulting channel is not exact. However, as in the diagonal case, it defines the associated incoherent LCU sampler and satisfies the probability domination bound \(\widetilde{p}_x\ge p_x/\Gamma_U\) for any \(x\).

As an application of the general construction, we now consider the fully connected XY-mixer, which can be used, for instance, to preserve cardinality constraints in QAOA \cite{hadfield_2019_xymixer,wang_2020_xymixer}. 
Let
\begin{eqnarray}
J_\mu &=& \sum_{i=1}^n \sigma_i^\mu,
\qquad \mu\in\{x,y,z\}, \\
H_{XY} &=& J_x^2+J_y^2,
\end{eqnarray}
and define
\[
U_{XY}(\beta) = e^{-i\beta H_{XY}}.
\]
Note that several conventions for the fully connected XY-mixer appear in
the literature. Our definition differs from the common convention
$\sum_{i<j}(X_iX_j+Y_iY_j)$ only by an overall factor and an additive
identity term, since
\[
J_x^2+J_y^2
=
2nI+2\sum_{i<j}(X_iX_j+Y_iY_j).
\]
The additive identity contributes only a global phase, while the overall factor can be absorbed into the variational angle $\beta$, hence, these differences are irrelevant for the present analysis.

In Appendix~\ref{appendix:xy_mixer}, we show that the XY-mixer acts on the spin-\(j\) block as
\[
U_{XY}(\beta)\big|_j=A_j(\beta),
\]
where
\[
A_j(\beta)
=
\sum_{m=-j}^{j}
e^{-i4\beta\left(j(j+1)-m^2\right)}
|j,m\rangle\langle j,m|.
\]
Thus, despite being highly non-local in the computational basis, the XY-mixer becomes particularly simple in the Schur--Weyl decomposition: it is block-diagonal and, within each block, fully diagonal. 

Substituting \(U_j=A_j(\beta)\) into the general permutation-invariant formula gives the coefficient function
\[
a_\beta(g)
=
\sum_{j\in\mathcal J_n}
(2j+1)\,
\operatorname{Tr}
\left[
A_j(\beta)D^j(g)^\dagger
\right].
\]
Then, as derived in Appendix~\ref{appendix:xy_mixer}, the coefficient function becomes
\begin{eqnarray}
&& a_\beta(\alpha,\vartheta,\chi) = 
\\ &&
\sum_{j\in\mathcal J_n}
(2j+1)
\sum_{m=-j}^{j}
e^{-i4\beta(j(j+1)-m^2)}
e^{im(\alpha+\chi)}
d^j_{mm}(\vartheta),
\end{eqnarray}
where \(d^j_{m m}(\vartheta)\) is the Wigner (small) \(d\)-matrix~\cite{Sakurai_1994_book}. 
The exact LCU is then
\[
U_{XY}(\beta)
=
\int
a_\beta(\alpha,\vartheta,\chi)
R(\alpha,\vartheta,\chi)^{\otimes n}
\frac{\sin\vartheta}{16\pi^2}
d\alpha\,d\vartheta\,d\chi,
\]
with
\[
\alpha\in[0,2\pi),\qquad
\vartheta\in[0,\pi],\qquad
\chi\in[0,4\pi),
\]
where we have used the standard Haar measure in Euler angles~\cite{leaser_su2_fourier_2012}.
The corresponding QPD cost $\Gamma_{XY}$ is again bounded by
\[
\Gamma_{XY}(\beta) \le \frac{(n+1)(n+2)(n+3)}{6}.
\]
This is remarkable because $H_{XY}$ is defined as a sum of non-commuting terms. Therefore, implementing $U_{XY}(\beta)$ typically requires approximations such as Trotterization or other more advanced Hamiltonian simulation methods, along with all-to-all connectivity.
In contrast, the LCU construction presented here achieves an exact implementation without Trotter error. This comes at the cost of a sampling overhead quantified by $\Gamma_{XY}$, but requires only a star connectivity in the exact setting, or no connectivity in the ancilla-free variant.

\section{Fourier-based LCU and QAOA \label{sec:qaoa_lcu}}

The Fourier-based LCU constructions developed in this work are a natural fit for QAOA and its alternating-operator variants \cite{farhi_2014_qaoa, hadfield_2019_xymixer, wang_2020_xymixer}.  
QAOA prepares a parametrized family of states by alternating problem-dependent phase-separation unitaries and mixing unitaries.  
Given a cost Hamiltonian $H_C$, a mixer Hamiltonian $H_M$, and an initial state $\ket{\psi_0}$, a depth-$p$ QAOA circuit prepares  
\begin{equation}
\ket{\psi(\beta,\gamma)}
=
\prod_{j=1}^p
e^{-i\beta_j H_M}e^{-i\gamma_j H_C}
\ket{\psi_0},
\end{equation}
where $\beta, \gamma \in \mathbb{R}^p$.  
The initial state is often the uniform superposition, although problem-informed warm-starts can also be used \cite{Egger_2021}.  
The angles are usually optimized in a classical outer loop to minimize or maximize  
\begin{equation}
\bra{\psi(\beta, \gamma)}
H_C
\ket{\psi(\beta, \gamma)}.
\end{equation}  
Each sampled bitstring corresponds to a candidate solution of the underlying combinatorial optimization problem.

Suppose that one or more phase-separation or mixing unitaries are replaced by Fourier-based LCU decompositions.  
If the quantum device is used only to generate bitstrings, we can use the ancilla-free LCU variant introduced above.
This approach avoids implementing the full coherent circuit, at the cost of increased sampling. Importantly, it still preserves the sample-quality guarantees established earlier. 
This leads to three natural optimization modes:

\begin{enumerate}
    \item \emph{Optimize the coherent target circuit.}  
    One may first optimize the fully coherent QAOA circuit using, e.g., classical (approximate) evaluations such as tensor-network or matrix-product-state simulation, Pauli-propagation, or analytic formulas available in special cases such as $p=1$ \cite{farhi_2014_qaoa, streif2019trainingquantumapproximateoptimization, rudolph2025paulipropagationcomputationalframework}.
    For some instance families, optimized angles can also be reused or transferred across instances, reducing or eliminating instance-specific training \cite{brandao2018fixedcontrolparametersquantum, Akshay_2021,  galda2023similaritybasedparametertransferabilityquantum}.

    \item \emph{Optimize or evaluate the ancilla-free LCU circuits directly.}  
    One may instead work with the ensemble of circuits arising from the ancilla-free LCU decompositions of Secs.~\ref{sec:fourier_diagonal} and~\ref{sec:fourier-nondiagonal}.  
    This can be combined with approximate classical evaluation, since these circuits are often simpler.

    \item \emph{Use a single LCU basis circuit as a variational ansatz.}  
    Finally, one may select a parametrized Fourier-based LCU basis circuit and jointly optimize its internal parameters $\vartheta$ together with the QAOA angles.  
    Denoting the resulting state by  
    $\ket{\psi_{\mathrm{LCU}}(\beta, \gamma, \vartheta)}$, one optimizes
    \begin{equation}    
    \bra{\psi_{\mathrm{LCU}}(\beta, \gamma, \vartheta)}
    H_C
    \ket{\psi_{\mathrm{LCU}}(\beta, \gamma, \vartheta)},
    \end{equation}  
    or other objective functions such as CVaR \cite{barkoutsos_2020_improving_variational_quantum_using_cvar, Barron_2024}.
\end{enumerate}

The third mode is conceptually different from ancilla-free LCU sampling.
When the Fourier-basis angles are freely optimized, the circuit is no longer a randomized approximation of the original unitary, but instead defines a new LCU-motivated variational ansatz. Consider, however, linear sample-based objective functions of the form $\mathbb{E}_{(j, X_j)}[f(X_j)]$, $f: \{0, 1\}^n \rightarrow \mathbb{R}$, where evaluation entails first sampling a branch $j$ from the LCU distribution $q_j$, and then sampling $X_j$ by measuring the corresponding quantum circuit.
This includes, for instance, the probability of sampling feasible high-quality bitstrings. 
In this setting, the performance of the randomized LCU samples is simply a weighted average over all branches.
Therefore, at least one branch must perform at least as well as this average. 
As a result, assuming global optimization and an infinite-shot limit, the performance of the optimized single-branch ansatz will match or exceed that of the randomized LCU sampler, while retaining its sampling overhead guarantees for this type of objective.
However, this argument does not apply to non-linear objectives such as CVaR. In such cases, the single-branch approach is only a heuristic approximation.

For diagonal Fourier-based LCU factors encoding constraint penalty terms, the third mode has a natural interpretation in terms of Lagrangian relaxation approaches \cite{gabbassov_2024_lagrangian,le_2024_lagrangian,sharma_2025_lagrangian}.  
Consider the constrained maximization problem  
\begin{eqnarray}
f^\star
=
&\max_{x\in\{0,1\}^n}& f(x)\\
&\text{subject to:}& g(x) = 0,
\end{eqnarray}
for functions $f: \{0, 1\}^n \rightarrow \mathbb{R}$ and $g: \{0, 1\}^n \rightarrow \{0, \ldots, m\}$.
The penalty-based approach is to optimize the unconstrained objective  
\begin{equation}
f_\lambda(x)=f(x)-\lambda g(x)^2,
\end{equation}
for a fixed penalty factor $\lambda > 0$.  
Suppose we can coherently implement diagonal unitaries corresponding to $f(x)$ and $g(x)$ and use a Fourier-based LCU to realize $\lambda g(x)^2$.
This yields diagonal LCU basis circuits corresponding to $f(x) + \theta_j g(x)$ for $j \in \{0, \ldots, m\}$.
The results of Sec.~\ref{sec:lcu} then imply that an ancilla-free Fourier-based LCU implementation can reproduce the bitstring-sampling behavior of the corresponding coherent penalty-QAOA circuit up to the stated sampling overhead.

The corresponding Lagrangian relaxation is  
\begin{equation}
d^\star
=
\min_{\mu\in\mathbb{R}}
\max_{x\in\{0,1\}^n}
\left\{ f(x)+\mu g(x)\right\},
\end{equation}
with weak duality implying $f^\star\le d^\star$.  
A QAOA phase separator for $f(x) + \mu g(x)$ promotes the Lagrange multiplier $\mu$ to a trainable circuit parameter.  
The resulting parametrized circuit therefore coincides with the basis circuit obtained from the LCU decomposition.
Consequently, for linear sample-based objectives as discussed before, the performance guarantees of the LCU approach extend to the Lagrangian relaxation circuit, provided the objective and parameter optimization are aligned.
For general variational objectives, this remains an LCU-motivated ansatz that may nevertheless perform very well in practice, but does not directly inherit the distributional guarantees of the LCU sampler.

For non-diagonal operations, such as Hamming-weight-preserving XY-mixers \cite{hadfield_2019_xymixer, wang_2020_xymixer}, the same interpretation no longer reduces to a classical Lagrangian.  
In this case, jointly optimizing the Fourier-based LCU parameters can be viewed as a quantum analogue of the Lagrangian relaxation: the additional parameters modify how the circuit explores the feasible subspace, rather than simply reweighting a classical constraint.  
This extends the Lagrangian perspective without a direct classical counterpart.

The main trade-off is between circuit complexity and sampling overhead.  
Replacing coherent unitaries by ancilla-free Fourier-based LCU circuits can reduce implementation requirements, but the associated sampling overhead multiplies across LCU-replaced factors and may grow exponentially with the QAOA depth $p$ in the worst case. Using a single LCU basis circuit as a variational ansatz can mitigate this effect while maintaining the theoretical guarantees for linear sample-quality objectives, such as the probability of sampling good bitstrings. 

When QAOA is used as a sampler of good diverse bitstrings, for example, in quantum approximate multi-objective optimization \cite{Kotil_2025} or sample-based quantum diagonalization \cite{Robledo_Moreno_2025, yu2025quantumcentricalgorithmsamplebasedkrylov, piccinelli2026quantumchemistryprovableconvergence}, it may be preferable to sample the LCU branch variables from their original Fourier-based LCU distribution and optimize only the coherent QAOA parameters instead of using a single LCU basis circuit as an ansatz.  
This better preserves the intended output distribution while avoiding overfitting to a single Fourier-basis circuit.

\section{Experimental Results\label{sec:demonstrations}}

In this section, we demonstrate Fourier-based LCU for QAOA on the densest $k$-subgraph problem \cite{feige_2001_dense_k_subgraph}, comparing penalty-based and mixer-based approaches for enforcing cardinality constraints. Additional applications are discussed in Appendix~\ref{sec:examples}.

Given a graph $\mathcal{G} = (\mathcal{V}, \mathcal{E}, w)$ with nodes $\mathcal{V} = \{1, \ldots, n\}$, edges $\mathcal{E} \subset \mathcal{V} \times \mathcal{V}$, and edge weights $w_{ij} \in \mathbb{R}$ for $(i, j) \in \mathcal{E}$, the densest $k$-subgraph problem is
\begin{eqnarray}
    \max_{x \in \{0, 1\}^n} && \sum_{(i, j) \in \mathcal{E}} w_{ij} x_i x_j\\
    \text{subject to:}&& \sum_{i=1}^n x_i = k,
\end{eqnarray}
where $k \leq n$ specifies the desired subgraph size.

We consider two approaches to enforce the cardinality constraint. First, we use a diagonal Fourier-based LCU to incorporate the constraint via a quadratic penalty term.
Second, we employ a non-diagonal Fourier-based LCU to implement a constraint-preserving XY-mixer.
In all experiments, we study QAOA with depth $p=1$ and warm-start from the initial state
\begin{eqnarray}
    \ket{\psi_0^k} &=& \bigotimes_{i=1}^n \left( \sqrt{1 - k/n}\ket{0} + \sqrt{k/n} \ket{1}\right),
\end{eqnarray}
which can be prepared using single-qubit $R_Y$ rotations with angle $\theta_{\text{init}} = 2\sin^{-1}(\sqrt{k/n})$.
This permutation-invariant initial state is biased towards feasible solutions. For the penalty-based warm-started QAOA circuits, we use a corresponding single-qubit mixer adapted to this initial state~\cite{Egger_2021}.
A corresponding circuit is illustrated in Fig.~\ref{fig:penalty_lcu_circuit} in Sec.~\ref{sec:dks_12_penalty}.

The Hamming weight $k$ obtained by measuring $\ket{\psi_0^k}$ follows a binomial distribution.
Therefore, the probability of sampling a feasible solution with Hamming weight $k$ is
\[
p^0_{\text{feasible}} = 
\left(\begin{array}{c}
     n  \\
     k
\end{array}\right)
\left(\frac{k}{n}\right)^k \left(1 - \frac{k}{n} \right)^{n-k} = \Theta\left(\frac{1}{\sqrt{n}}\right),
\]
where the latter scaling follows from Stirling's approximation when $k/n$ is bounded away from $0$ and $1$.
Hence, $\ket{\psi_0^k}$ has inverse-polynomial squared overlap with the Dicke state $\ket{D_{n,k}}$, i.e., the uniform superposition of all computational basis states with Hamming weight $k$.

We start with two illustrative simulation studies on a 12-node random 3-regular graph, using exact simulations for both the penalty-based and XY-mixer-based approaches, with $k = \lfloor n/3 \rfloor = 4$.
We then consider a larger instance with $n=106$ nodes and $k = \lfloor n/3 \rfloor = 35$.
This graph is constructed from the heavy-hex connectivity of the target quantum device and densified using three layers of SWAP gates (see Appendix~\ref{sec:heavy_hex_graphs} for more details).
For this instance, we demonstrate the penalty-based approach at scale on a real quantum computer.
In all cases, we assume unweighted graphs, i.e., $w_{ij} = 1$ for all edges $(i, j) \in \calE$.
The optimal solutions are computed using CPLEX \cite{ibm_cplex}, yielding values of $4$ edges (out of $18$) for the 12-node graph and $98$ edges (out of $328$) for the 106-node graph.

\subsection{12-node densest $k$-subgraph with penalty term (simulation)\label{sec:dks_12_penalty}}

\begin{figure}
    \centering
    \includegraphics[width=1\linewidth]{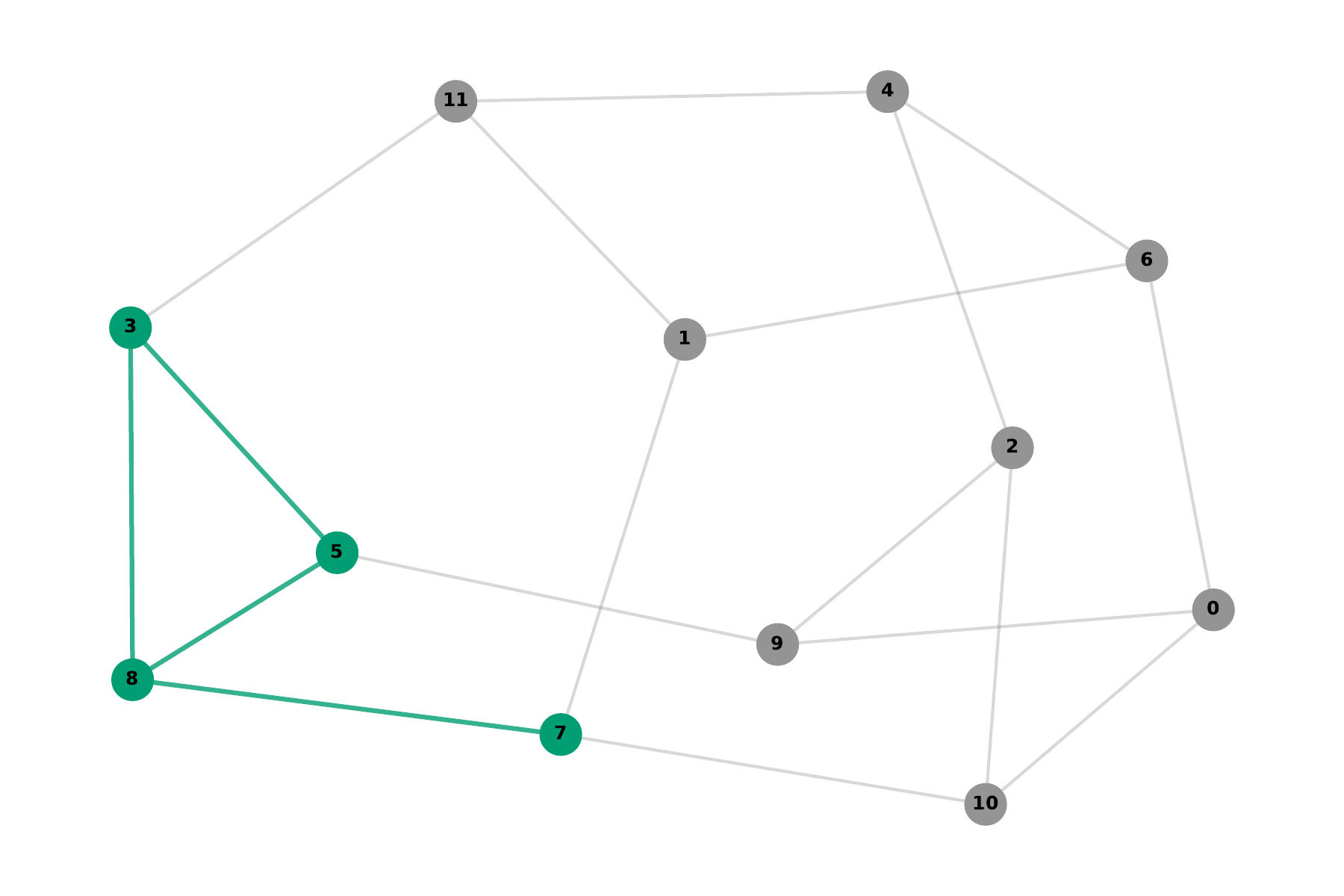}
    \caption{The considered 3-regular graph and corresponding optimal subgraph (highlighted in green) for $k=4$.}
    \label{fig:dks_12_solution}
\end{figure}

We consider a 12-node random 3-regular unweighted graph with target subgraph size $k=4$, and map the densest $k$-subgraph problem to a Quadratic Unconstrained Binary Optimization (QUBO) problem via a quadratic penalty term. Specifically, we solve
\begin{eqnarray}
    \max_{x \in \{0, 1\}^n} && \sum_{(i, j) \in \mathcal{E}} w_{ij} x_i x_j - \lambda ( \mathbf{1}^T x - k )^2,
\end{eqnarray}
where $\lambda > 0$ is a penalty factor.
We set
\[
\lambda = 1 + \max_{i \in \calV} \sum_{(i, j) \in \calE} |w_{ij}|,
\]
which guarantees that violating the cardinality constraint is never beneficial. Indeed, adding or removing a single node can change the objective by at most $(\lambda - 1)$, whereas changing the Hamming weight away from $k$ incurs a penalty of at least $\lambda$.
The considered graph and its corresponding solution are shown in Fig.~\ref{fig:dks_12_solution}.

\begin{figure}[t]
    \centering
    \resizebox{\columnwidth}{!}{%
\begin{quantikz}
    \lstick{$q_1$}
        & \gate{R_Y(\theta_{\mathrm{init}})}
        & \gate[wires=3]{e^{-i\gamma H_1}}
        & \gate{R_Z(\theta)}
        & \gate{R_Y(-\theta_{\mathrm{init}})}
        & \gate{R_Z(\beta)}
        & \gate{R_Y(\theta_{\mathrm{init}})}
        & \qw \\
        & \setwiretype{n}\push{\vdots}
        & \setwiretype{n}
        & \setwiretype{n}
        & \setwiretype{n}
        & \setwiretype{n}
        & \setwiretype{n}
        & \setwiretype{n}
        \\
    \lstick{$q_n$}
        & \gate{R_Y(\theta_{\mathrm{init}})}
        & \ghost{e^{-i\gamma H_1}}
        & \gate{R_Z(\theta)}
        & \gate{R_Y(-\theta_{\mathrm{init}})}
        & \gate{R_Z(\beta)}
        & \gate{R_Y(\theta_{\mathrm{init}})}
        & \qw
\end{quantikz}%
    }
    \caption{Penalty-based LCU basis circuit with a warm-started initial state and mixer \cite{Egger_2021}, the cost operator $H_1$ for the objective function corresponding to graph $\mathcal{G}$, and an LCU basis layer for the cardinality constraint composed solely of $R_Z$ gates.}
    \label{fig:penalty_lcu_circuit}
\end{figure}
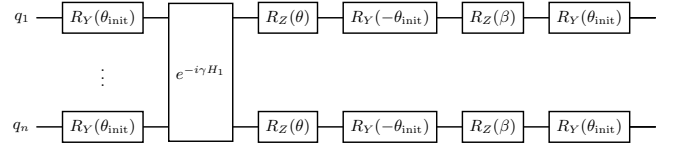

We denote by $H_1$ the Ising Hamiltonian corresponding to the objective function and by $H_2$ the Hamiltonian corresponding to the penalty term (including the penalty factor), such that $H = H_1 + H_2$ represents the full QUBO.
We compare a fully coherent implementation of the QAOA cost operator with an implementation where $H_1$ is implemented coherently and $H_2$ is realized via Fourier-based LCU.
This reduces the connectivity requirements from all-to-all interactions to those induced by the graph $\mathcal{G}$, which can be substantially less demanding. The corresponding LCU basis circuit is illustrated in Fig.~\ref{fig:penalty_lcu_circuit}.

We perform the following five experiments:
\begin{enumerate}
    \item Optimize the fully coherent QAOA circuit parameters to maximize $\brakets{H}$, yielding optimal parameters $(\beta^*, \gamma^*)$. Parameter optimization is performed via grid search followed by COBYLA \cite{powell_1994_cobyla} for fine-tuning.
    \item Evaluate the corresponding Fourier-based LCU approximation at $(\beta^*, \gamma^*)$ and compute the resulting sampling overhead $\Gamma$.
    \item Re-optimize $(\beta, \gamma)$ to maximize the $\text{CVaR}_{1/\Gamma}$ using weighted LCU samples collected from all $n+1$ circuits, noting that $\Gamma$ depends on $\gamma$. 
    \item Optimize a single LCU basis circuit, jointly maximizing over $(\beta, \gamma, \theta)$ using $\text{CVaR}_{1/\Gamma}$ as objective. For comparability, we fix $\Gamma$ to the value obtained in Experiment~3.
    \item Re-optimize the fully coherent QAOA circuit using $\text{CVaR}_{1/\Gamma}$ as the objective, again fixing $\Gamma$ to the value obtained in Experiment~3, to put the LCU-based results into context.
\end{enumerate}

For all experiments, we report the resulting expectation values, the sampling overhead $\Gamma$, the corresponding CVaR values where applicable, the probability of sampling a feasible solution, the probability of sampling an optimal feasible solution, and the expectation value conditioned on feasibility.

\begin{figure*}
    \centering
    \includegraphics[width=1\linewidth]{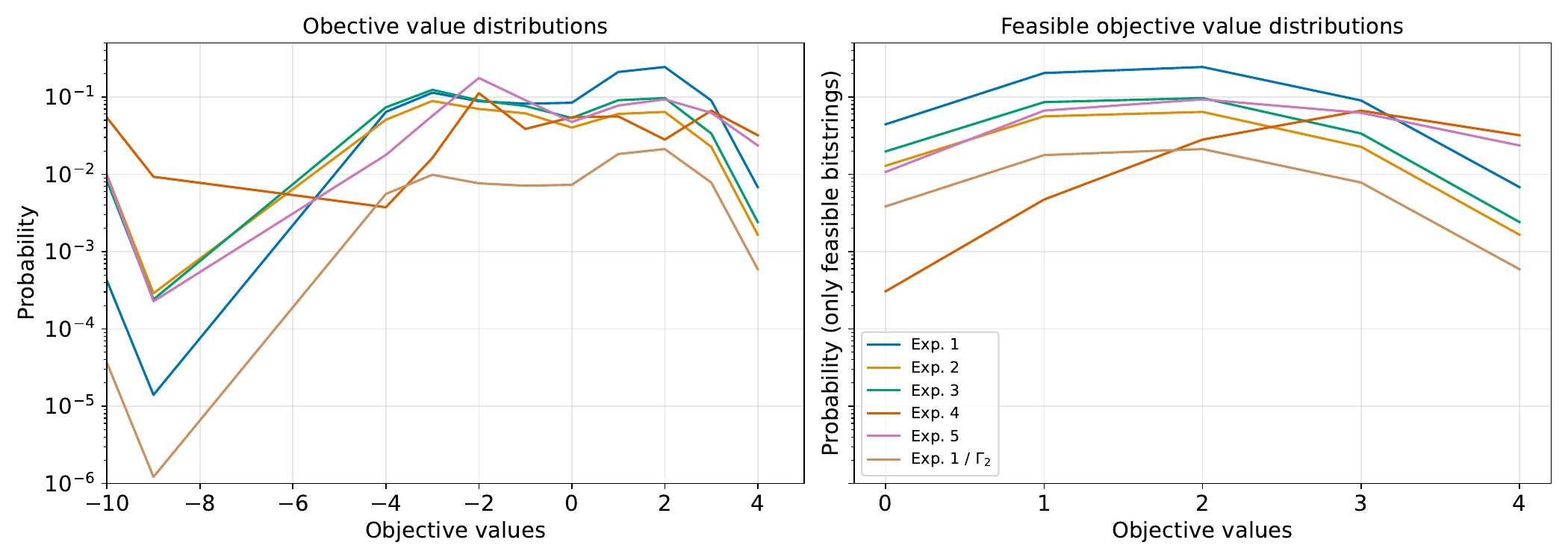}
    \caption{Results for the 12-node densest $k$-subgraph simulations with a penalty term. 
    (Left) Distribution of objective values obtained from samples of the different experiments, restricted to the range $[-10, 4]$ for clarity. 
    (Right) Corresponding distributions restricted to feasible samples (unnormalized). Both panels also show ``Exp.~1$/\Gamma_2$'', representing the coherent distribution scaled by the sampling overhead from Experiment~2, thereby illustrating the theoretical lower bound for Experiment~2.}
    \label{fig:dks_12_penalty}
\end{figure*}

The results are shown in Tab.~\ref{tab:dks_12_penalty} and Fig.~\ref{fig:dks_12_penalty}. As expected, the distributions obtained from Fourier-based LCU circuits using the same $(\beta^*, \gamma^*)$ are lower bounded by the coherent distributions scaled by $1/\Gamma$. Furthermore, the corresponding CVaR values provide upper bounds on the coherent $\brakets{H}$, although these bounds are relatively loose. 
Notably, optimizing a single LCU basis circuit yields the highest probability of sampling an optimal solution.

\begin{table*}
    \centering
    \begin{tabular}{clcccccc}
        \# & Experiment & $\brakets{H}$ & $\Gamma$ & $\text{CVaR}_{1/\Gamma}$ & $\mathbb{P}$-feasible & $\mathbb{P}$-optimal & $\brakets{H}_{\text{feasible}}$ \\
        \hline
        1 & Coherent QAOA            &  -0.3344 & --      & --     & 0.5898 & 0.0068 & 1.6796 \\
        2 & Fourier-based LCU              & -16.9382 & 11.5189 & 2.2987 & 0.1592 & 0.0016 & 1.6383 \\
        3 & Fourier-based LCU re-optimized &  -8.6698 & 12.5041 & 2.4818 & 0.2384 & 0.0024 & 1.6364 \\
        4 & Single LCU basis circuit & -24.2318 & 12.5041 & 3.3969 & 0.1325 & 0.0317 & 2.9645 \\
        5 & Coherent QAOA (CVaR)     &  -8.0497 & 12.5041 & 3.2950 & 0.2565 & 0.0236 & 2.0840
    \end{tabular}
    \caption{Results for the 12-node densest $k$-subgraph simulations with a penalty term: expectation values, sampling overhead $\Gamma$, CVaR values, probabilities of sampling feasible and optimal solutions, and expectation values conditioned on feasibility.}
    \label{tab:dks_12_penalty}
\end{table*}

\subsection{12-node densest $k$-subgraph with XY-mixer (simulation)}

We consider the same densest $k$-subgraph instance as in Sec.~\ref{sec:dks_12_penalty}.
Instead of enforcing the cardinality constraint via a penalty term, we combine the biased initial state with the cardinality-preserving XY-mixer discussed in Sec.~\ref{sec:fourier-nondiagonal}. It is known that QAOA with constraint-preserving mixers performs best when initialized in an extremal eigenstate of the mixer Hamiltonian in the feasible subspace \cite{hadfield_2019_xymixer, wang_2020_xymixer, He_2023}. In the present case, the relevant feasible subspace is defined by $\mathbf{1}^T x = k$, and the ideal initial state within this subspace would be the Dicke state $\ket{D_{n,k}}$.
Thus, the chosen initial state has squared overlap $|\brakets{D_{n,k}\mid \psi_0^k}|^2 = p^0_{\text{feasible}} = \Theta(1 / \sqrt{n})$ with $\ket{D_{n,k}}$. Since neither the cost operator nor the XY-mixer couples subspaces of different Hamming weight, the probability of sampling a feasible state is preserved throughout the coherent evolution. For $k/n$ bounded away from 0 and 1, this corresponds to a sampling overhead of only $\mathcal{O}(\sqrt{n})$ due to the Dicke state approximation. The XY-mixer is implemented using a second-order Trotter decomposition with five Trotter steps.

To facilitate a consistent comparison across XY-mixer experiments, we use $\text{CVaR}_{\eta}$ of the QUBO objective from Sec.~\ref{sec:dks_12_penalty} as the optimization objective, rather than post-selecting feasible samples. Since the probability of sampling feasible solutions is known in the fully coherent setting, we initialize the risk level at $\eta = p^0_{\text{feasible}}$ and, where appropriate, rescale it to $\eta = p^0_{\text{feasible}} / \Gamma$ to account for LCU sampling overhead. Empirically, we observe that CVaR optimization converges rapidly to the optimal objective value. To further refine the optimization, we include the probability of sampling an optimal solution as a secondary objective with weight $10^{-5}$.

\begin{figure}[t]
    \centering
    \resizebox{\columnwidth}{!}{%
\begin{quantikz}
    \lstick{$q_1$}
        & \gate{R_Y(\theta_{\mathrm{init}})}
        & \gate[wires=3]{e^{-i\gamma H_1}}
        & \gate{R_Z(\alpha)}
        & \gate{R_Y(\vartheta)}
        & \gate{R_Z(\chi)}
        & \qw \\
        & \setwiretype{n}\push{\vdots}
        & \setwiretype{n}
        & \setwiretype{n}
        & \setwiretype{n}
        & \setwiretype{n}
        & \setwiretype{n}
        \\
    \lstick{$q_n$}
        & \gate{R_Y(\theta_{\mathrm{init}})}
        & \ghost{e^{-i\gamma H_1}}
        & \gate{R_Z(\alpha)}
        & \gate{R_Y(\vartheta)}
        & \gate{R_Z(\chi)}
        & \qw
\end{quantikz}%
    }
    \caption{XY-mixer-based LCU basis circuit with a warm-started initial state and mixer, the cost operator $H_1$ for the objective function corresponding to graph $\mathcal{G}$, and an LCU basis layer for the cardinality constraint composed solely of $R_Z$ and $R_Y$ gates.}
    \label{fig:xy_lcu_circuit}
\end{figure}
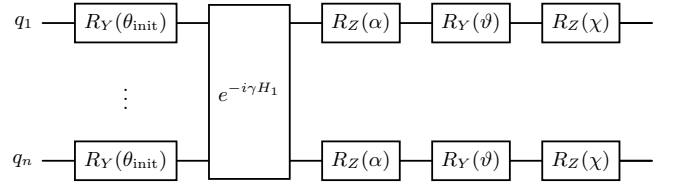

Since the LCU for the XY-mixer comprises a continuous family of basis circuits, explicit enumeration is infeasible and random sampling is required to approximate it.
We draw $10^6$ samples $(\alpha, \vartheta, \chi)$ from the Haar-uniform distribution to form a candidate pool, compute the corresponding LCU weights, and subsequently sample $1000$ circuits according to the resulting normalized discrete distribution. These sampled circuits are used for all LCU-based experiments. To estimate the sampling overhead $\Gamma$ for a given value of $\beta$, we draw $10^5$ additional samples. The corresponding LCU basis circuit is illustrated in Fig.~\ref{fig:xy_lcu_circuit}.

We run the following experiments similar to Sec.~\ref{sec:dks_12_penalty}:
\begin{enumerate}
    \item Optimize the fully coherent XY-QAOA circuit parameters to maximize $\text{CVaR}_{p^0_{\text{feasible}}}$ using the penalty-based QUBO objective function, yielding optimal parameters $(\beta^*, \gamma^*)$. As before, we first perform a grid search followed by fine-tuning with COBYLA.
    \item Evaluate the sampled LCU circuits at $(\beta^*, \gamma^*)$ and estimate the resulting sampling overhead $\Gamma$.
    \item Re-optimize $(\beta, \gamma)$ to maximize $\text{CVaR}_{p^0_{\text{feasible}}/\Gamma}$ using sampled solutions collected across all sampled LCU circuits. Note that $\Gamma$ depends on the current value of $\beta$. 
    \item Optimize a single parametrized LCU basis circuit over $(\gamma, \vartheta, \chi)$, again using $\text{CVaR}_{p^0_{\text{feasible}}/\Gamma}$ as objective and fixing $\Gamma$ to the value obtained in Experiment~3 for comparability. Here, $\alpha$ is irrelevant since the corresponding $R_Z$ rotation appears immediately before measurement, and $\beta$ is absent as the mixer Hamiltonian is realized via the LCU decomposition.
    \item Re-optimize the coherent QAOA circuit using $\text{CVaR}_{p^0_{\text{feasible}}/\Gamma}$ with the same $\Gamma$ as in Experiment~3 for comparison.
\end{enumerate}
Additionally, we run the following experiments to compare penalty-based and XY-mixer-based constraint handling:
\begin{enumerate}
\setcounter{enumi}{5}
    \item Optimize the XY-QAOA circuit parameters to maximize $\text{CVaR}_{1/\Gamma_3^p}$, where $\Gamma_3^p$ denotes the sampling overhead from the penalty-based Experiment~3 in Sec.~\ref{sec:dks_12_penalty}. Apart from the choice of objective, this setup matches Experiment~1.
    \item Optimize a single parametrized LCU basis circuit over $(\beta, \gamma, \alpha, \vartheta, \chi)$ using $\text{CVaR}_{1/\Gamma_3^p}$ as the objective, mirroring Experiment~4 with the objective of Experiment~6.
\end{enumerate}

For all experiments, we report expectation values, sampling overhead $\Gamma$, risk levels $\eta$ and corresponding CVaR values, the probability of sampling a feasible solution, the probability of sampling a feasible optimal solution, and the expectation value when conditioning on feasible solutions.

\begin{figure*}
    \centering
    \includegraphics[width=1\linewidth]{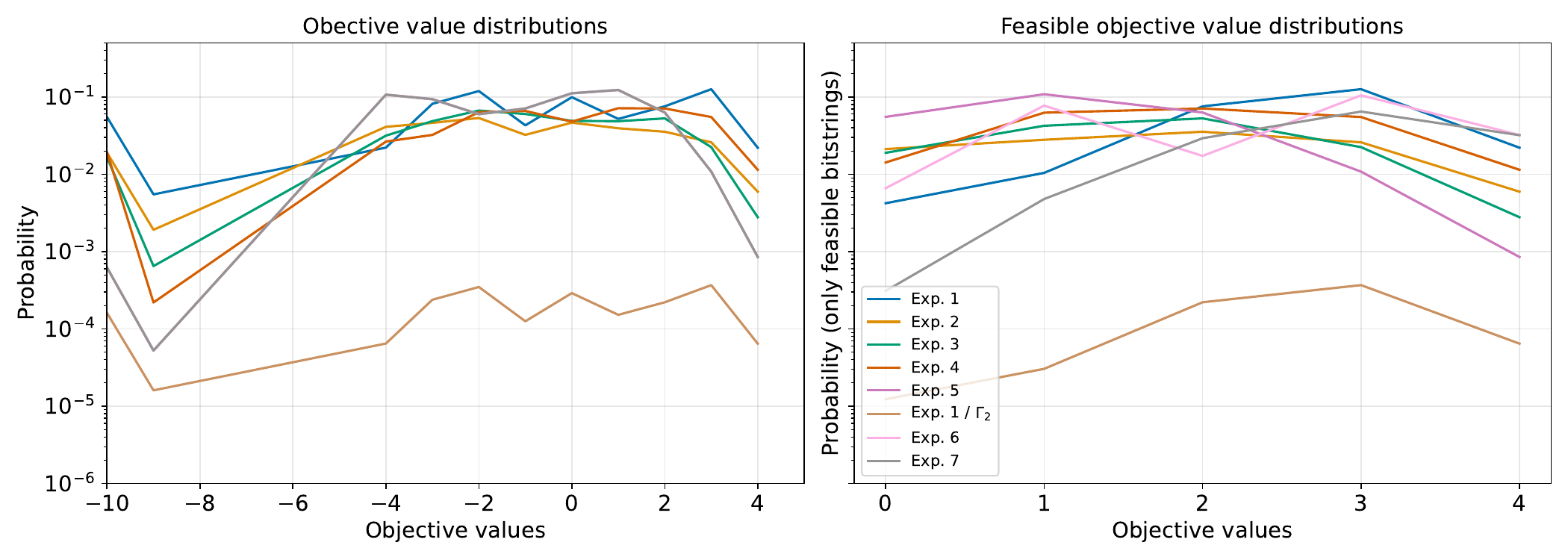}
    \caption{Results for the 12-node densest $k$-subgraph simulations with XY-mixer. 
    (Left) Distribution of objective values obtained from samples of the different experiments, restricted to the range $[-10, 4]$ for clarity. 
    (Right) Corresponding distributions restricted to feasible samples (unnormalized). Both panels also show ``Exp.~1$/\Gamma_2$'', representing the distribution of Experiment~1 scaled by the sampling overhead of Experiment~2, thereby illustrating the theoretical lower bound for Experiment~2.}
    \label{fig:dks_12_xy}
\end{figure*}

The results are summarized in Tab.~\ref{tab:dks_12_xy} and Fig.~\ref{fig:dks_12_xy}. 
As predicted, the probabilities obtained from the Fourier-based LCU circuits (Experiment~2) are lower bounded by the coherent probabilities scaled by $1/\Gamma$. Moreover, the corresponding CVaR results provide upper bounds on the coherent results of Experiment~1, often attaining the optimal value. Overall, the $\Gamma$-based bounds (brown line in Fig.~\ref{fig:dks_12_xy}) are quite loose, as approximate simulations closely track the fully coherent results in practice. This suggests that the effective sampling overhead for larger problem instances may be substantially smaller than the worst-case theoretical guarantees. Large values of $\Gamma$ are required only to satisfy the bound for a small subset of probabilities, where it can indeed be tight.

Comparing the penalty-based and XY-mixer-based approaches at identical CVaR levels, we consistently observe superior performance for the XY-mixer. This is expected, in particular, for the single LCU basis circuits, as the XY-mixer basis circuit naturally generalizes the penalty-based basis circuit.
Notably, a single XY-mixer LCU basis circuit achieves nearly the same probability of sampling an optimal solution as the fully coherent Trotterized XY-QAOA circuit and, in fact, a slightly higher CVaR value.
These results indicate that, for the problem considered, training a single LCU basis circuit with a CVaR objective offers a competitive---and significantly more hardware-friendly---alternative to fully coherent XY-QAOA implementations. 

\begin{table*}
    \centering
    \begin{tabular}{clccccccc}
        \# & Experiment & $\brakets{H}$ & $\Gamma$ & $\eta$ & $\text{CVaR}_{\eta}$ & $\mathbb{P}$-feasible & $\mathbb{P}$-optimal & $\brakets{H}_{\text{feasible}}$ \\
        \hline
        1 & Coherent XY-QAOA (CVaR)   &  -7.8079 & --       & 0.2384 & 2.6525 & 0.2384 & 0.0220 & 2.6350 \\
        2 & Fourier-based LCU               & -42.5502 & 342.1372 & 0.0007 & 4.0000 & 0.1159 & 0.0059 & 1.7146 \\
        3 & Fourier-based LCU re-optimized  & -27.2421 & 336.9378 & 0.0007 & 4.0000 & 0.1401 & 0.0030 & 1.6245 \\
        4 & Single LCU basis circuit  & -16.8427 & 336.9378 & 0.0007 & 4.0000 & 0.2143 & 0.0114 & 1.9390 \\
        5 & Coherent XY-QAOA (CVaR)   &  -8.6014 & 336.9378 & 0.0007 & 4.0000 & 0.2384 & 0.0017 & 1.5866 \\
        \hline
        6 & Coherent XY-QAOA ($\text{CVaR}_{1/\Gamma_3^p}$) &  -8.1167 &  12.5041 & 0.0800 & 2.3600 & 0.2384 & 0.0323 & 2.3322 \\
        7 & Single LCU basis circuit ($\text{CVaR}_{1/\Gamma_3^p}$) & -23.9291 &  12.5041 & 0.0800 & 3.4006 & 0.1312 & 0.0320 & 2.9411
    \end{tabular}
    \caption{Results for the 12-node densest $k$-subgraph simulations with XY-mixer: expectation values, sampling overhead $\Gamma$, CVaR risk level $\eta$, resulting CVaR values, probabilities of sampling feasible and optimal solutions, and expectation values conditioned on feasibility.}
    \label{tab:dks_12_xy}
\end{table*}

\subsection{106-node densest $k$-subgraph (hardware) \label{sec:106_dks}}

For the 106-node graph, statevector simulations are no longer feasible. For the penalty-based approach, the expectation value of the fully coherent warm-started QAOA at depth $p=1$ can be evaluated analytically, since the mixer consists only of single-qubit gates. This is not possible for the XY-mixer, which acts globally. We therefore use the analytic evaluation of the warm-started penalty-based circuit to obtain good circuit parameters $(\beta_1^*, \gamma_1^*)$ together with the corresponding analytic expectation value $\brakets{H} = 23.5452$ as a reference.

The considered circuits follow the structures shown in Figs.~\ref{fig:penalty_lcu_circuit} and \ref{fig:xy_lcu_circuit}. They use 106 qubits and 886 CZ gates to implement $H_1$ and the SWAP gates, with a two-qubit-gate depth of 25. This corresponds to approximately 35 CZ gates in parallel, i.e., on average acting on about 70 of the 106 qubits at a time.

We run the following experiments on the IBM Quantum system \emph{ibm\_boston} via the IBM Quantum platform \cite{ibm_quantum_platform}, sampling $2^{15} = 32768$ shots per circuit and repeating the full series of experiments ten times for robustness:
\begin{enumerate}
    \item Evaluate all $107$ Fourier-based LCU circuits for the penalty-based approach, aggregate the sampled probabilities using the LCU weights, and compute the $\text{CVaR}_{1/\Gamma}$ for comparison with $\brakets{H}$, where $\Gamma = 104.1328$.
    
    \item Optimize a single penalty-based LCU basis circuit by jointly maximizing $\text{CVaR}_{1/\Gamma}$ over $(\beta, \gamma, \theta)$, with the same $\Gamma$ as in Experiment~1, using COBYLA and initializing at $(\beta_1^*, \gamma_1^*, \theta_{j^*})$, where $\theta_{j^*}$ denotes the LCU basis angle achieving the best CVaR value in Experiment~1. The resulting angles are denoted $(\beta_2^*, \gamma_2^*, \theta_2^*)$.
    
    \item Optimize a single XY-mixer-based LCU basis circuit by maximizing $\text{CVaR}_{1/\Gamma}$ over $(\gamma, \vartheta, \chi)$, using the same $\Gamma$ as in Experiments~1 and 2. The initial parameters are chosen as $(\gamma_2^*, \vartheta_{\beta_2^*, \theta_2^*, \theta_{\text{init}}}, \chi_{\beta_1^*, \theta_2^*, \theta_{\text{init}}})$, such that they reproduce the optimal solution from Experiment~2 up to a global phase (see~Appendix~\ref{sec:single_xy_circuit_initialization}). This ensures that the optimization starts from the best solution found in Experiment~2 and exploits the additional degrees of freedom to further improve the results. As before, $\beta$ is absorbed into the LCU parameters, while $\alpha$ can be neglected for $p=1$ as it controls an $R_Z$ rotation immediately before measurement. 
\end{enumerate}

\begin{figure}
    \centering
    \includegraphics[width=1\linewidth]{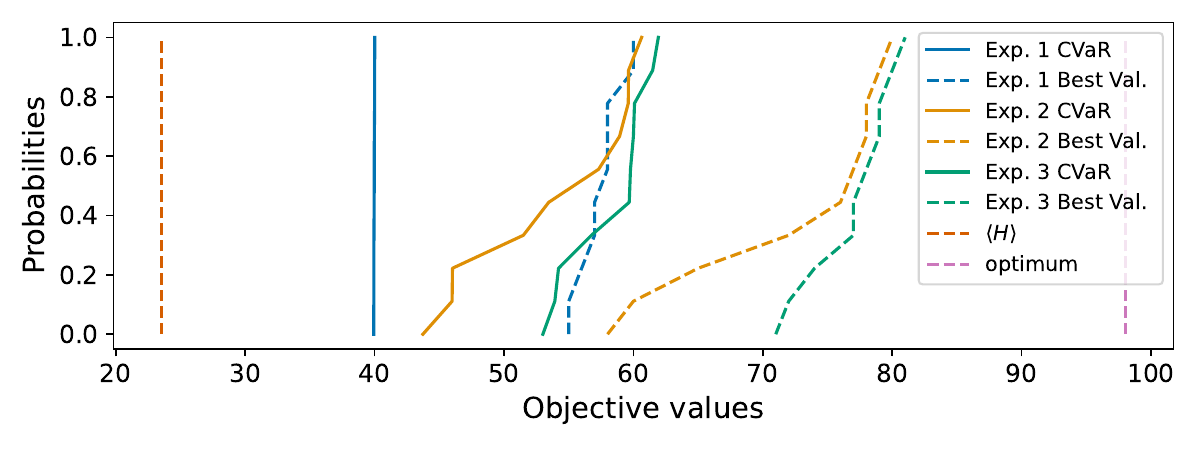}
    \caption{Cumulative distribution functions of the CVaR values and best found solutions over ten repetitions for each of the three experiments on the 106-node SWAP-extended heavy-hex graph. The analytic expectation value of coherent penalty-QAOA and the optimal solution computed by CPLEX are also shown for reference.}
    \label{fig:dks_106_distributions}
\end{figure}

For all experiments, we report the maximum and average over the ten repetitions of the best solution found, the final CVaR value, and the final probability of sampling feasible solutions. The aggregated results are shown in Tab.~\ref{tab:dks_106_aggregated} and Fig.~\ref{fig:dks_106_distributions}. The 106-node problem graph and the best solution found on the quantum computer are shown in Fig.~\ref{fig:dks_106_best}, while the full results and a graph showing an optimal solution are provided in Appendix~\ref{sec:dks_106_additional_results}.
Note that the true analytic sampling overhead in Experiments~2 and 3 would naturally differ from $\Gamma = 104.1328$ as the circuit parameters are updated. We intentionally fix $\Gamma$ to the same value across all experiments to maintain a consistent CVaR evaluation metric, enabling a fair comparison of the outcomes. This distinction is particularly relevant for the XY-mixer in Experiment~3, where the theoretical sampling overhead would be substantially larger. Nevertheless, our hardware results suggest that the single-branch LCU ansatz can perform exceptionally well even with significantly fewer samples, indicating that the heuristic approach can remain effective well below the worst-case theoretical bounds.

The results show a consistent improvement in solution quality across the experiments. The probability of sampling feasible solutions from the initial state $\ket{\psi_0^k}$ is $8.22\%$, as computed from the binomial distribution. Assuming that the penalty-based coherent implementation would lead to a similar probability, and noting that the XY-mixer would preserve this probability exactly, the achieved probabilities between 2.72\% and 4.87\% indicate that the practical overhead is much smaller than the worst-case overhead $\Gamma = 104.1328$. 

\begin{table*}
\begin{tabular}{cl|cc|cc|cc}
&& \multicolumn{2}{c|}{Best solution} & \multicolumn{2}{c|}{$\text{CVaR}_{1/\Gamma}$} & \multicolumn{2}{c}{$\mathbb{P}$-feasible} \\
\hline
\# & Experiment & max. & avg. & max. & avg. & max. & avg. \\
\hline
1 & Fourier-based LCU & 60 & 57.4 & 40.0142 & 39.9707 & 0.0275 & 0.0272 \\
2 & Single penalty-based LCU basis circuit  & 80 & 72.3 & 60.6369 & 53.6794 & 0.0762 & 0.0487 \\
3 & Single XY-mixer-based LCU basis circuit & 81 & 76.8 & 61.9263 & 58.0851 & 0.0533 & 0.0468
\end{tabular}
\caption{Aggregated results over ten repetitions for the three experiments on the 106-node SWAP-extended heavy-hex graph. Reported are the maximum and average (across repetitions) of best found solutions throughout the optimization, the final $\text{CVaR}_{1/\Gamma}$ values for $\Gamma = 104.1328$, and the final probabilities of sampling feasible solutions.}
\label{tab:dks_106_aggregated}
\end{table*}

\begin{figure}
    \centering
    \includegraphics[width=\linewidth]{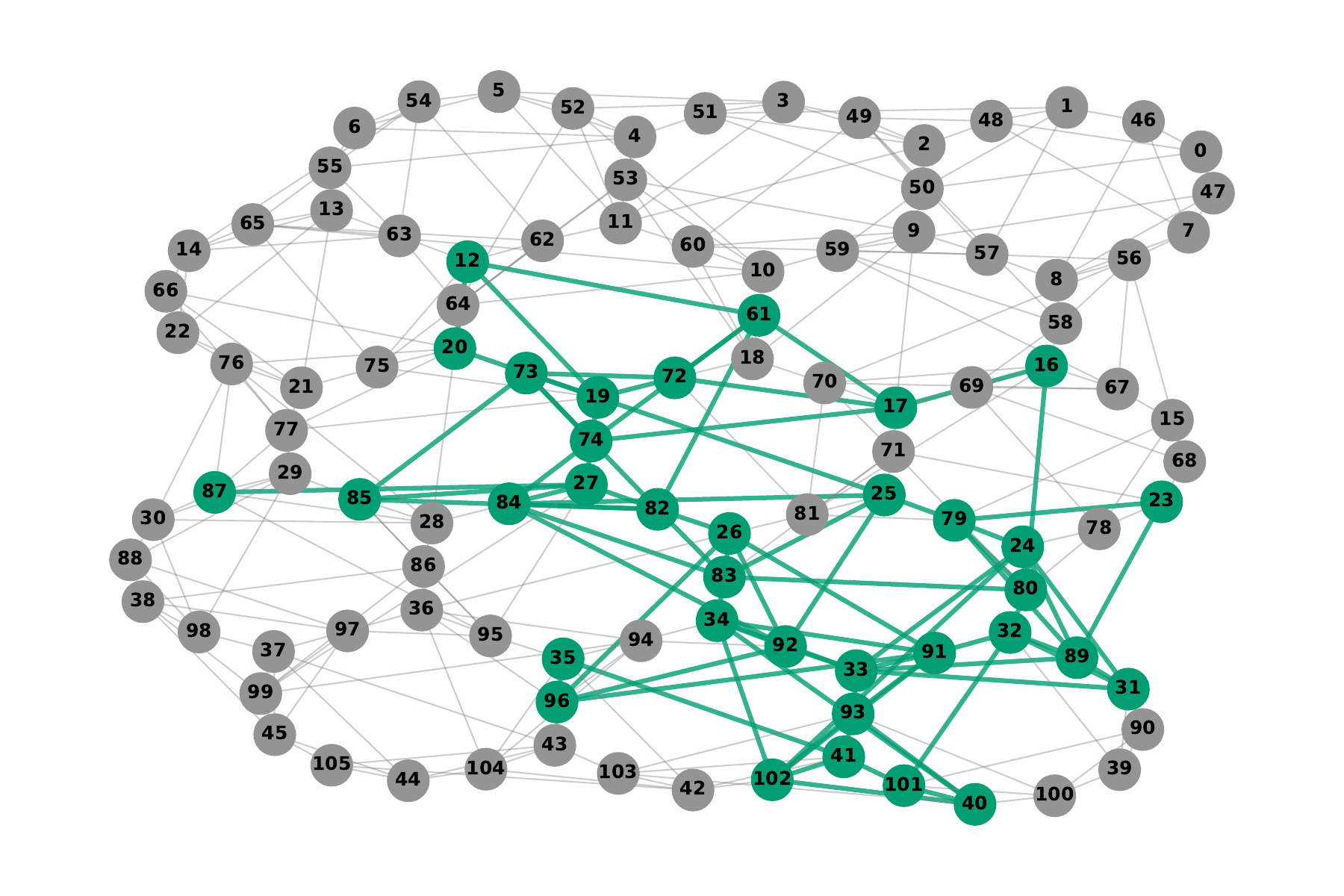}
    \caption{Problem graph with 106 nodes resulting from a $5 \times 3$ heavy-hex grid extended by three SWAP layers. The green nodes and edges indicate the best found solution on the quantum computer.}
    \label{fig:dks_106_best}
\end{figure}

\section{Conclusion \& Outlook\label{sec:conclusion}}

In this work, we introduced a Fourier-based LCU framework to represent complex quantum circuits---such as those arising in quantum optimization---as weighted combinations of simpler circuits. The approach provides a trade-off between circuit complexity and sampling overhead, thereby extending the reach of applications that can be studied on today's quantum computers.

We developed the construction for diagonal unitaries and extended it to non-diagonal permutation-invariant unitaries, such as the XY-mixer. In the context of QAOA, this allows all-to-all penalty and mixer operations to be replaced by collections of substantially simpler circuits, often involving only single-qubit gates, while retaining provable sample-quality guarantees up to polynomial sampling overhead. We also highlighted connections between coherent penalty formulations and Lagrangian relaxations, suggesting a broader perspective on constraint handling in quantum optimization.

Our numerical and hardware experiments on the densest $k$-subgraph problem demonstrate that this trade-off can be useful in practice. Our results show that Fourier-based LCU decompositions can be applied to instances with more than one hundred qubits, and that the theoretical bounds can be relatively loose and that practical costs can be substantially lower. While we focused on cardinality equality constraints, the construction extends to many other relevant classes of constraints, including inequalities.

Several directions remain open. First, it would be valuable to identify even more efficient---possibly approximate---LCU representations for broader families of unitaries, as well as to better understand the limitations. Second, beyond optimization, the same ideas may apply to other sampling-based quantum algorithms, such as sampling-based quantum diagonalization, whenever samples can be evaluated classically and exact sampling from the target distribution is unnecessary. Third, Fourier-based LCU may enable modular decompositions of problems that are locally separable except for a global coupling term representable by LCU, allowing large instances to be executed through smaller circuits combined by a global quasi-probabilistic reconstruction.
Finally, parameter optimization remains an important challenge. For LCUs implementing a coherent circuit, parameters optimized for the original circuit---broadly studied, for example in QAOA---can be reused. However, optimizing extended variational ansaetze with additional degrees of freedom may be more difficult and requires further study. 

Fourier-based LCU therefore provides a systematic route to studying algorithms and problem instances whose coherent implementations are beyond current hardware capabilities. By replacing complex coherent operations with weighted combinations of simpler circuits, it broadens the range of quantum optimization experiments that can be run today and offers a flexible design principle for more hardware-compatible quantum algorithms.

\subsection*{Code availability}
The source code supporting the results of this work is publicly available~\cite{fourier_lcu_2026_repo}.

\subsection*{Acknowledgments}
We thank David Sutter, Lukas Schmitt, Victor Martinez, Shashanka Ubaru, and Ken Clarkson for insightful and constructive technical discussions.

IBM Bob and GPT‑5.5 were used to assist with code development and text editing during the preparation of this manuscript. All scientific ideas, analyses, and conclusions are those of the authors, and all AI-assisted code and text were carefully reviewed, manually checked, and validated prior to inclusion.

\appendix

\section{Hilbert space decomposition\label{appendix:hilbert_decomposition}}
The general Schur--Weyl duality gives~\cite{bacon_2007_schur_transform}
\begin{equation}
    (\mathbb{C}^d)^{\otimes n} \cong \bigoplus_{\lambda\in \mathcal{I}_{d,n}} \mathcal{Q}^d_\lambda \otimes \mathcal{P}_\lambda,
\end{equation}
where $\mathcal{I}_{d,n}=\{\lambda=(\lambda_1,\ldots,\lambda_d)\mid\lambda_1\geq\ldots\geq\lambda_d\geq 0, \quad \sum_{i=1}^d\lambda_i=n\}$ denotes the set of partitions of $n$ with at most $d$ parts. $\mathcal{Q}^d_\lambda$ and $\mathcal{P}_\lambda$ are irreducible representations of $\mathcal{U}_d$ and $\mathcal{S}_n$, respectively.

We now specialize to the case $d=2$.
In this setting, partitions with at most two parts can be written as $\lambda=(n-k,k)$ for $k=0,\ldots,\lfloor n/2 \rfloor$.
Thus, $\mathcal{I}_{2,n}$ is naturally identified with $\mathcal{J}_n$ via the bijection $(n-k,k) \mapsto j = \tfrac{n}{2} - k$.

From~\cite[Theorem~2.14]{zhang_2014_schur_weyl}, the dimensions are
\begin{align*}
    \dim(\mathcal{Q}^2_\lambda)&=n-2k+1=2j+1,\\ 
    m_j := \dim(\mathcal{P}_\lambda)&=\binom{n}{\frac{n}{2}-j}-\binom{n}{\frac{n}{2}-j-1},
\end{align*}
where the latter follows from the hook-length formula~\cite{Frame_Robinson_Thrall_1954}.

Substituting these expressions back into the Schur--Weyl decomposition yields the form
\begin{equation}
    (\mathbb C^2)^{\otimes n}
\simeq
\bigoplus_{j\in\mathcal J_n}
\mathbb C^{2j+1}\otimes \mathbb C^{m_j},
\end{equation}
where the label $j$ indexes the spin sectors, and $m_j$ denotes their multiplicities.

\section{Continuous LCU in $SU(2)$\label{appendix:fourier_in_groups}}
We seek a continuous LCU decomposition of the form
\[
U
=
\int_{SU(2)}
a_U(g)\, R(g)^{\otimes n}\, d\mu(g),
\]
where \(d\mu(g)\) is the normalized Haar measure on \(SU(2)\), and \(a_U(g)\in\mathbb{C}\).
The following derivation is based on the theory of compact Lie groups. We refer the reader to Ref.~\cite{leaser_su2_fourier_2012} for a more detailed treatment.

To determine \(a_U(g)\), we use Fourier analysis on \(SU(2)\).
By the Peter--Weyl theorem, the matrix elements of the irreducible representations \(D^j(g)\) form an orthogonal basis for functions on \(SU(2)\). In particular, any integrable function \(f\) admits Fourier coefficients
\[
\widehat{f}(j)
=
\int_{SU(2)}
f(g)\, D^j(g)^\dagger\, d\mu(g),
\]
and can be reconstructed via the Fourier inversion formula
\[
f(g)
=
\sum_{j}
(2j+1)\,
\operatorname{Tr}\!\left[
\widehat{f}(j)\, D^j(g)
\right].
\]

Comparing with the block decomposition
\[
U=
\bigoplus_{j\in\mathcal J_n}
U_j\otimes I_{m_j},
\]
we interpret the matrices \(U_j\) as the Fourier coefficients associated with each irreducible representation. The corresponding coefficient function is therefore given by the inverse Fourier transform
\[
a_U(g)
:=
\sum_{j\in\mathcal J_n}
(2j+1)\,
\operatorname{Tr}\!\left[
U_j\, D^j(g)^\dagger
\right].
\]

Substituting this expression into the integral and using the orthogonality relations of the irreducible representations yields
\begin{align}
\int_{SU(2)}
a&_U(g)\, R(g)^{\otimes n}\, d\mu(g)\\
&=
\bigoplus_{j\in\mathcal J_n}
\left(
\int_{SU(2)}
a_U(g)\, D^j(g)\, d\mu(g)
\right)\otimes I_{m_j}\\
&=
\bigoplus_{j\in\mathcal J_n}
U_j\otimes I_{m_j}
=
U.
\end{align}

\section{Bounding $\Gamma_U$}\label{appendix:bound_on_gamma}
Here we derive a general bound on the cost $\Gamma_U$ as introduced in Sec.~\ref{sec:fourier-nondiagonal}.
Recall that
\[
\Gamma_U = \alpha_U^2,
\qquad
\alpha_U = \int_{SU(2)} |a_U(g)|\, d\mu(g).
\]

Applying the Cauchy--Schwarz inequality to the functions \(|a_U(g)|\) and \(1\), and using the fact that $d\mu$ is a normalized Haar measure, we obtain
\begin{equation*}
\left(\int_{SU(2)} |a_U(g)|\, d\mu(g)\right)^2\le \int_{SU(2)} |a_U(g)|^2\, d\mu(g).
\end{equation*}

Substituting the definition of $a_U(g)$, we obtain
\begin{align*}
&\int |a_U(g)|^2\, d\mu(g)\\
&\begin{split}
=&\sum_{j,k}
(2j+1)(2k+1)\\
&\int
\operatorname{Tr}\!\left[
U_j D^j(g)^\dagger
\right]
\operatorname{Tr}\!\left[
D^k(g) U_k^\dagger
\right]
\, d\mu(g).
\end{split}
\end{align*}

Using the identity \(\operatorname{Tr}(A B)=\sum_{m,n} A_{mn} B_{nm}\), we rewrite the expression as
\begin{align*}
&\sum_{j,k}
(2j+1)(2k+1)
\sum_{m,n,m',n'}
(U_j)_{mn}
(U_k^\dagger)_{n'm'} \\
&\qquad \times
\int
\overline{D^j_{mn}(g)}\,
D^k_{m'n'}(g)
\, d\mu(g).
\end{align*}

We now use the Schur orthogonality relations of irreducible representations~\cite{leaser_su2_fourier_2012}:
\[
\int_{SU(2)}
\overline{D^j_{mn}(g)}\,
D^k_{m'n'}(g)
\, d\mu(g)
=
\frac{1}{2j+1}
\delta_{jk}\delta_{mm'}\delta_{nn'}.
\]

Applying this identity collapses the sums:
\begin{align*}
\int |a_U(g)|^2\, d\mu(g)
&=
\sum_{j}
(2j+1)
\sum_{m,n}
(U_j)_{mn}
(U_j^\dagger)_{nm}.
\end{align*}

Recognizing the Frobenius norm, we obtain
\[
\int |a_U(g)|^2\, d\mu(g)
=
\sum_{j\in\mathcal J_n}
(2j+1)\|U_j\|_F^2=
\sum_{j\in\mathcal J_n}
(2j+1)^2,
\]
where we used that $U_j$ is unitary in the last equality.
Therefore
\[
\Gamma_U
\le
\sum_{j\in\mathcal J_n}(2j+1)^2
=
\frac{(n+1)(n+2)(n+3)}{6}
=
O(n^3).
\]

\section{LCU for the XY-mixer\label{appendix:xy_mixer}}
Let
\[
S_\mu:=\frac{J_\mu}{2}\quad\text{and}\quad S^2:=S_x^2+S_y^2+S_z^2.
\]
Then
\[
J_x^2+J_y^2
=
4(S_x^2+S_y^2)
=
4(S^2-S_z^2),
\]
where $S^2$ and $S_z^2$ commute.
Moreover, on each spin-\(j\) sector, \(S^2\) has eigenvalue \(j(j+1)\), while \(S_z\) has eigenvalues
\[
m=-j,-j+1,\dots,j.
\]
These properties are standard results from the theory of angular momentum. A more detailed discussion can be found, for example, in Ref.~\cite{Sakurai_1994_book}.

Therefore, the XY-mixer is diagonal in the standard $\ket{j, m}$ basis of each irreducible subspace:
\[
U_{XY}(\beta)\big|_j=A_j(\beta),
\]
where
\[
A_j(\beta)
=
\sum_{m=-j}^{j}
e^{-i4\beta\left(j(j+1)-m^2\right)}
|j,m\rangle\langle j,m|.
\]

After substituting the above into the general expression for permutation-invariant unitaries, we obtain the coefficient function
\[
a_\beta(g)
=
\sum_{j\in\mathcal J_n}
(2j+1)\,
\operatorname{Tr}
\left[
A_j(\beta)D^j(g)^\dagger
\right].
\]
To make this expression more explicit, we parametrize using Euler angles and use the standard decomposition of irreducible representation matrices~\cite{Sakurai_1994_book}
\[
D^j_{m m'}(\alpha,\vartheta,\chi)
=
e^{-im\alpha}
d^j_{m m'}(\vartheta)
e^{-im'\chi},
\]
where \(d^j_{m m'}(\vartheta)\) is the Wigner (small) \(d\)-matrix. 
Since \(A_j(\beta)\) is diagonal, only diagonal matrix elements contribute to the trace, yielding
\begin{eqnarray}
&& a_\beta(\alpha,\vartheta,\chi)
\\ 
&=&
\sum_{j\in\mathcal J_n}
(2j+1)
\sum_{m=-j}^{j}
e^{-i4\beta(j(j+1)-m^2)}
D^j_{mm}(\alpha, \vartheta, \chi)^*\\
&=&
\sum_{j\in\mathcal J_n}
(2j+1)
\sum_{m=-j}^{j}
e^{-i4\beta(j(j+1)-m^2)}
e^{im(\alpha+\chi)}
d^j_{mm}(\vartheta).
\end{eqnarray}

\section{Further Applications of Efficient Fourier LCUs}
\label{sec:examples}

We now present several optimization problems that particularly benefit from Fourier-based LCUs, either by reducing circuit depth, relaxing connectivity requirements, or enabling efficient implementations of otherwise costly penalty terms.

\subsection{Graph problems on Erd\H{o}s-R\'enyi graphs with high edge probability}

We first consider graph optimization problems such as MAXCUT or Maximum Independent Set (MIS) on unweighted Erd\H{o}s-R\'enyi graphs. An Erd\H{o}s-R\'enyi graph on \(n\) nodes is constructed by including each edge \((i,j)\) with probability \(p \in [0,1]\).

When \(p > 1/2\), it is often advantageous to begin with an implementation of the cost operator corresponding to the complete graph. Using Fourier-based LCU, this can be realized with sampling overhead \(n+1\). The edges that are not present in the target graph are then coherently removed. As a result, the required connectivity is effectively shifted from the original graph to its complement, which is typically much sparser in this regime.

\subsection{Skewed Sherrington-Kirkpatrick model}

The Sherrington-Kirkpatrick (SK) model corresponds to MAXCUT on a complete graph with random edge weights, commonly drawn from \(U(\{-1,+1\})\). In the \emph{skewed} SK model, we assign edge weights as follows: with probability \(p\), the weight is set to \(+1\), and with probability \(1-p\), it is set to \(-p/(1-p)\), ensuring zero mean edge weight.

A QAOA cost operator for this model can be implemented by first realizing the complete graph with all edge weights set to \(+1\) using a Fourier-based LCU with sampling overhead \(n+1\). The edges with negative weights are then implemented coherently by correcting for the excess \(+1\) contribution. This reduces the all-to-all connectivity requirement to only those edges with negative weights. For \(p > 1/2\), this yields a substantial advantage over standard circuit constructions.

\subsection{Maximum Independent Set with indicator-function penalties}

Let \(\mathcal{G} = (\mathcal{V}, \mathcal{E})\) be a graph with vertex set \(\mathcal{V} = \{1,\ldots,n\}\) and edge set \(\mathcal{E} \subseteq \mathcal{V} \times \mathcal{V}\). The MIS problem can be formulated as
\begin{eqnarray}
    \max_{x \in \{0,1\}^n} && \sum_{i=1}^n x_i \\
    \text{subject to:} && x_i x_j = 0 \quad \forall (i,j) \in \mathcal{E}.
\end{eqnarray}
In standard QAOA formulations, violations are penalized using the quadratic function
\[
g(x) = \sum_{(i,j)\in \mathcal{E}} x_i x_j .
\]

An alternative approach proposed in the literature \cite{Bucher_2025} replaces this quadratic penalty with an indicator function \(h\), defined as \(h(x)=0\) for feasible solutions and \(h(x)=1\) otherwise. If the quadratic penalty \(g\) can be implemented coherently, our Fourier-based LCU construction immediately yields an efficient decomposition of the indicator penalty with sampling overhead bounded by \(\Gamma \leq 1+|\mathcal{E}|\).

\subsection{Index tracking with risk constraints}

We finally consider index tracking with risk constraints for a portfolio of \(n\) assets, characterized by a covariance matrix \(\Sigma\) of returns. The optimization problem is given by
\begin{eqnarray}
    \min_{x \in \{0,1\}^n} && (x - \mathbf{1})^T \Sigma (x - \mathbf{1}) \\
    \text{subject to:} && \ell \leq x^T \Sigma x \leq u .
\end{eqnarray}
Here, the objective minimizes the tracking error relative to the full index, while the constraint enforces acceptable risk levels. We assume \(u < \mathbf{1}^T \Sigma \mathbf{1}\), as otherwise the trivial solution \(x=\mathbf{1}\) is optimal.

Provided that \(\Sigma\) is sufficiently sparse to allow an efficient QAOA implementation of the cost operator, and that its entries can be rounded so that $x^T \Sigma x$ takes only a moderate number of distinct values, the Fourier-based LCU framework enables efficient penalty constructions for the risk constraints. This avoids the need to add slack variables and increase the polynomial degree to quartic, which would result in significantly more complex and resource-intensive circuit implementations.

\section{Construction of SWAP-extended heavy-hex graphs\label{sec:heavy_hex_graphs}}

In this section, we describe the construction of the SWAP-extended heavy-hex graphs used for the 106-node densest $k$-subgraph problem in Sec.~\ref{sec:106_dks}.
We start with a heavy hexagonal lattice of $5 \times 3$ heavy hex cells. This graph admits a 3-edge-coloring, partitioning all edges into three disjoint color classes such that no two edges sharing a vertex have the same color. This coloring enables parallel operations on all edges within a single color class without conflicts.

To extend the connectivity beyond the base heavy-hex topology, SWAP layers are systematically added to create long-range connections between qubits. For each SWAP layer, all edges from one of the three color classes are selected (cycling through the colors sequentially), and SWAP gates are applied to these edges in parallel. The new qubit connections induced by these SWAP gates are then permanently added to the graph topology. The number of SWAP layers is a tunable parameter controlling the degree of long-range connectivity, with each layer contributing the edges corresponding to one round of parallel SWAP gates on a specific color class of the coupling map.
In this work, we use three SWAP layers and the resulting connectivity is illustrated in Figs.~\ref{fig:dks_106_best} and \ref{fig:dks_106_opt}.

\section{Initialization of single XY-mixer-based LCU circuits\label{sec:single_xy_circuit_initialization}}

In Experiment~3, discussed in Sec.~\ref{sec:106_dks}, we initialize the circuit parameters as 
$(\gamma_2^*, \vartheta_{\beta_2^*, \theta_2^*, \theta_{\text{init}}}, \chi_{\beta_1^*, \theta_2^*, \theta_{\text{init}}})$ such that they reproduce the optimal solution from Experiment~2 up to a global phase. This ensures that Experiment~3 starts from the best solution found in Experiment~2.

After optimization, the LCU gates and the warm-started mixer in Experiment~2 are given on each qubit by $R_Z(\theta_2^*) R_Y(-\theta_{\text{init}}) R_Z(\beta_2^*) R_Y(\theta_{\text{init}})$. In contrast, each single-qubit component of the XY-mixer LCU basis circuit has the form $R_Z(\alpha) R_Y(\vartheta) R_Z(\chi)$. 
Since the last $R_Z$ gate acts immediately before a measurement, it can be omitted.
We can then choose $\vartheta$ and $\chi$ such that the remaining unitary is equivalent, up to a global phase, to the warm-started mixer for the parameters $\theta_{\text{init}}, \beta_2^*$ and $\theta_2^*$.

\section{Additional results for the 106-node densest $k$-subgraph problem\label{sec:dks_106_additional_results}}

\begin{figure}
    \centering
    \includegraphics[width=\linewidth]{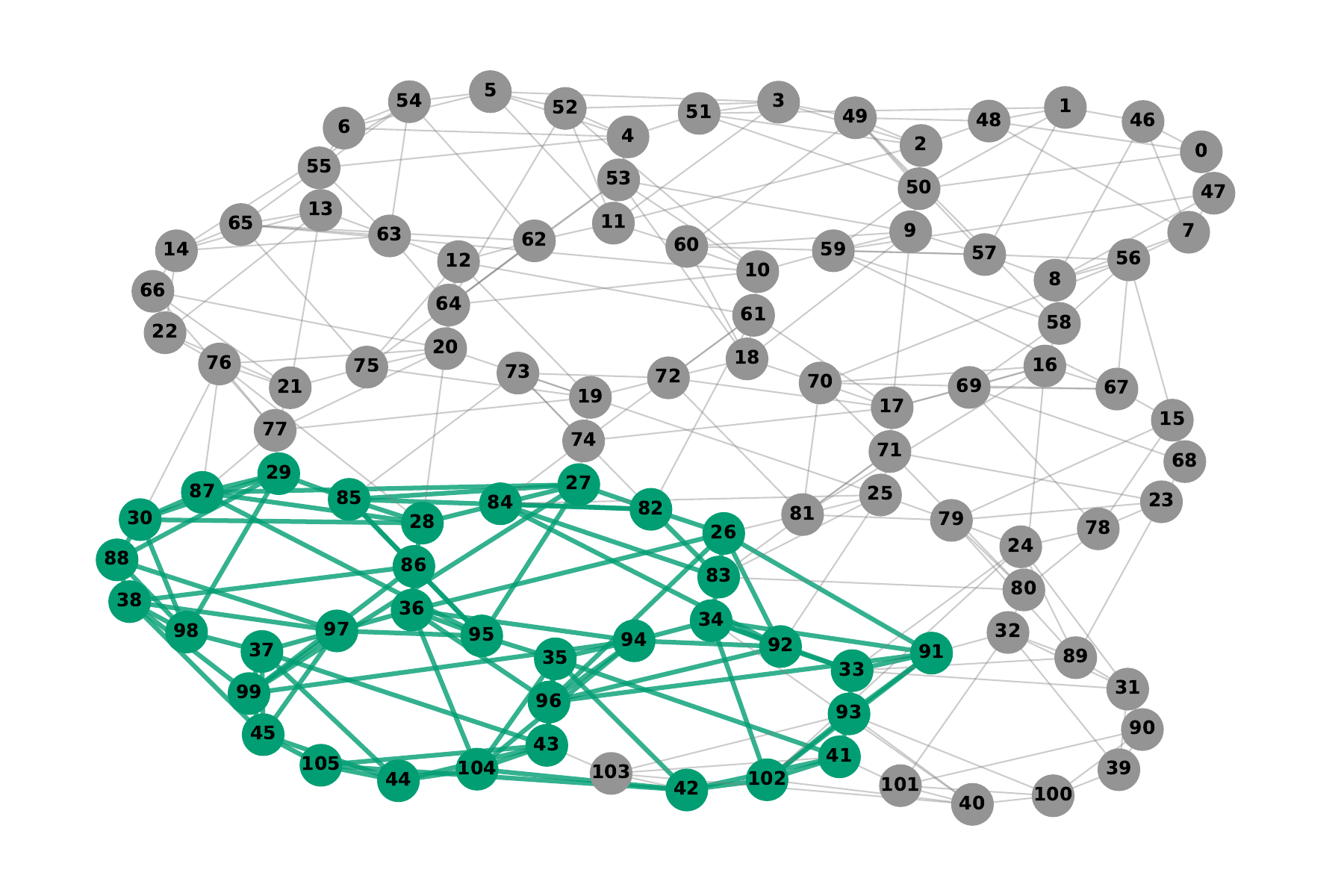}
    \caption{Problem graph with 106 nodes obtained from a $5 \times 3$ heavy-hex grid extended by three SWAP layers. The green nodes and edges show an optimal solution found by CPLEX.}
    \label{fig:dks_106_opt}
\end{figure}

This section presents additional results for the 106-node densest $k$-subgraph problem introduced in Sec.~\ref{sec:106_dks}. Tab.~\ref{tab:dks_106_complete_results} reports, for each of the three experiments across ten repetitions, the best solutions found during optimization, final CVaR values, and final probabilities of feasible solutions. Fig.~\ref{fig:dks_106_opt} shows the problem graph as well as an optimal solution.

\begin{table*}
\centering
\begin{tabular}{ccccc}
\# & Experiment & Best solution & $\text{CVaR}_{1/\Gamma}$ & $\mathbb{P}$-feasible \\
\hline
\multirow{11}{*}{1} & \multirow{11}{*}{\rotatebox[origin=c]{90}{Fourier-based LCU}} & 55 & 39.979368 & 0.027335 \\
& & 55 & 40.014231 & 0.027151 \\
& & 56 & 39.971362 & 0.027403 \\
& & 57 & 39.967406 & 0.027162 \\
& & 57 & 40.003091 & 0.027470 \\
& & 58 & 39.944893 & 0.027174 \\
& & 58 & 39.950036 & 0.026828 \\
& & 58 & 40.003486 & 0.027094 \\
& & 60 & 39.932215 & 0.027251 \\
& & 60 & 39.940416 & 0.027059 \\
\hline
\multirow{11}{*}{2}& \multirow{11}{*}{\rotatebox[origin=c]{90}{\shortstack{Single penalty-based \\ LCU basis circuit}}} & 58 & 43.731586 & 0.076233 \\
& & 60 & 46.017811 & 0.036224 \\
& & 65 & 45.989811 & 0.042755 \\
& & 72 & 51.495139 & 0.052856 \\
& & 76 & 57.310259 & 0.050476 \\
& & 77 & 59.621373 & 0.043213 \\
& & 78 & 53.457805 & 0.045746 \\
& & 78 & 58.930699 & 0.047607 \\
& & 79 & 60.636929 & 0.047485 \\
& & 80 & 59.602707 & 0.043915 \\
\hline
\multirow{11}{*}{3}& \multirow{11}{*}{\rotatebox[origin=c]{90}{\shortstack{Single XY-mixer-based \\ LCU basis circuit}}} & 71 & 53.002255 & 0.046448 \\
& & 72 & 53.927588 & 0.040436 \\
& & 74 & 54.213813 & 0.053284 \\
& & 77 & 56.756475 & 0.046631 \\
& & 77 & 59.999144 & 0.049316 \\
& & 78 & 59.677374 & 0.047699 \\
& & 79 & 60.086256 & 0.048737 \\
& & 79 & 61.926266 & 0.047028 \\
& & 80 & 61.487594 & 0.047028 \\
& & 81 & 59.773820 & 0.041504
\end{tabular}
\caption{Complete results over ten repetitions for the three experiments on the 106-node SWAP-extended heavy-hex graph. Reported are the best found solutions throughout the optimization, the final $\text{CVaR}_{1/\Gamma}$ values for $\Gamma = 104.1328$, and the final probabilities of sampling feasible solutions. Rows are sorted by the best found objective within each experiment.
}
\label{tab:dks_106_complete_results}
\end{table*}

\bibliography{references}

\end{document}